\documentclass[11pt]{revtex4}
\usepackage[colorlinks=true,linkcolor=red,citecolor=blue]{hyperref}
\usepackage{amsfonts}
\usepackage{amsmath}
\usepackage{amssymb}
\usepackage{epsfig}
\usepackage{graphicx}
\usepackage{color}
\begin{document}
\title{Analysis of Two-body Charmed $B$ Meson Decays in Factorization-Assisted Topological-Amplitude Approach}
\author{Si-Hong Zhou$^1$, Yan-Bing Wei$^1$, Qin Qin$^1$, Ying Li$^{2,4}$, Fu-Sheng Yu$^{3}$, Cai-Dian L\"u$^{1,4}$}
\affiliation{1.~Institute of High Energy Physics, Beijing 100049, People's Republic of China;\\
2.~Department of Physics, Yantai University, Yantai 264005,People's Republic of China\\
3.~School of Nuclear Science and Technology, Lanzhou University, Lanzhou 730000, People's Republic of China\\
4.~State Key Laboratory of Theoretical Physics, Institute of Theoretical Physics, Chinese Academy of Sciences, Beijing 100190, People's Republic of China}
\date{\today}
\begin{abstract}
Within the factorization-assisted topological-amplitude  approach, we  study the two-body charmed $B$ meson decays $B_{u,d,s} \to D^{(*)}M$, with $M$ denoting a light pseudoscalar (or vector) meson. The meson decay constants and transition form factors are factorized out from the hadronic matrix element of topological diagrams. Therefore the effect of SU(3) symmetry breaking is retained, which is different from the conventional topological diagram approach. The number of free nonperturbative parameters to be fitted from experimental data is also much less. Only four universal nonperturbative parameters $\chi^C$, $\phi^C$, $\chi^E$ and $\phi^E$ are introduced to describe the contribution of the color suppressed tree and $W$-exchanged diagrams for all the decay channels. With the fitted parameters from 31 decay modes induced by $b\to c$ transition, we then predict the branching fractions of 120 decay modes induced by both $b\to c$ and $b\to u$ transitions. Our results are well consistent with the measured data or to be tested in the LHCb and Belle-II experiments in the future. Besides, the SU(3) symmetry breaking, isospin violation and $CP$ asymmetry are also investigated.
\end{abstract}
\maketitle
\section{Introduction}\label{sec:1}

Due to the large mass and fast weak decay property of the top quark, $B$ mesons are the only weakly decaying mesons containing quarks of the third generation. Their nonleptonic weak decays provide direct access to the parameters of the Cabibbo-Kobayashi-Maskawa (CKM) matrix and to the study of $CP$ violation (for reviews, see, for examples \cite{Cheng:2009xz,Wang:2014sba}). Simultaneously, the studies of these decays can also provide some insight into the long distance non-perturbative structure of QCD as well as some hints of the new physics beyond the standard model (SM). To achieve these goals, the BaBar and Belle experiments at the $e^+e^-$ $B$-factories \cite{Bevan:2014iga} and the LHCb experiment \cite{LHCb-implications} at the Large Hadron Collider (LHC) have already performed high precision measurements of nonleptonic weak decays. In the era of the Belle-II \cite{Abe:2010gxa} and LHCb upgrade \cite{LHCb-implications}, the experimental analysis will be pushed towards new frontiers of precision.

In particular, the direct $CP$ violation in a decay process requires at least two contributing amplitudes with different weak and strong phases. In the SM, the weak phases can be accommodated in the CKM matrix, while no satisfactory first-principle calculations can yield the strong phases till now. To study the information of strong phases from the  non-leptonic $B$ decays is a tough work. The basic theoretical framework for the  non-leptonic $B$ decays is based on the operator product expansion and renormalization group equation, which allow us to write the amplitude of a decay $\overline B\to f$ generally as follows:
\begin{equation}
\mathcal{A} (\overline B\to f)= \langle f|\mathcal{H}_{eff}|\overline B\rangle= {G_{F}\over \sqrt 2} V_{CKM}\sum_i C_i(\mu)\langle f|O_i(\mu)|\overline B\rangle,
\end{equation}
where $\mathcal{H}_{eff}$ is the effective weak Hamiltonian, with $O_i(\mu)$ denoting the relevant local four-quark operators, which govern the decays in question. The CKM factors $V_{CKM}$ and the Wilson coefficients $C_i$ describe the strength with which a given operator enters the Hamiltonian. Now the only challenge for theorists is how to calculate the matrix elements $\langle f|O_i(\mu)|\overline B\rangle$ in QCD reliably. For decades we have applied the ``factorization"  hypothesis to estimate the matrix element of the four-quark operators through the product of the matrix elements of the corresponding quark currents. In the 1980s, the ``color transparency" viewpoints \cite{Dugan:1990de, Wirbel:1985ji, Neubert:1997uc} were used to justify this concept, while it could be put on a rigorous theoretical basis in the heavy-quark limit for a variety of $B$ decays about ten years ago \cite{Beneke:1999br, Ali:2007ff, Bauer:2001cu}. Alternatively, another useful approach is provided by the decomposition of their amplitudes in terms of different decay topologies and to apply the SU(3) flavor symmetry of strong interactions to derive relations between them \cite{Buras:1998ra}. Supplemented by isospin symmetry, the approximate SU(3) flavor symmetry and various ``plausible" dynamical assumptions, the diagrammatic approach has been used extensively for non-leptonic $B$ decays \cite{Chau:1990ay}.

Among $B$ decays, the charmed hadronic $B$ mesons decays $B\to D^{(*)} M$, where $M$ is a light meson, are of great interest for several reasons. Firstly, due to the existence of charm quark, the charmed hadronic decay processes have no contribution from penguin operators, so theoretical uncertainties involved in the relevant QCD dynamics become much less. Secondly, for the $b \to c$ transiting processes, since the CKM factors are real, the phases associated with these decay amplitudes afford us the information of clean strong interactions. Thirdly, for some typical decays such as $\overline B_s^0\to D_s^{(*)^\pm}K^\mp$ and $\overline B_d^0\to D^{(*)^\pm}\pi^\mp$, both $b \to c$ and $b \to u$ transitions contribute to their amplitudes, the interferences between which will allow us to extract the CKM phase $\gamma$ effectively \cite{gamma}. Lastly, these processes serve as a good testing ground for various theoretical issues in hadronic $B$ decays, such as factorization hypothesis, SU(3) symmetry breaking, and isospin violation. Experimentally, plenty of two-body charmed hadronic $B$ decays have been observed from the heavy flavor experiments, such as Belle, BaBar, D0, CDF and LHCb \cite{Amhis:2014hma}. Besides the available data, many new modes are being measured in LHCb. In the theoretical side, much attention has already been paid to these charmed hadronic $B$ decays. The color-favored decays $B\to D^{(*)} \pi$ were firstly explored in the framework of the factorization hypothesis \cite{Wirbel:1985ji, Neubert:1997uc}. Including the next-leading order corrections of vertexes, the factorization of this kind of processes has been proved within the QCD factorization approach \cite{Beneke:1999br} and the soft-collinear effective theory \cite{Bauer:2001cu}, which implies the final-state interactions of these decays are small. However, the color suppressed modes $B^0 \to \bar D^0 \pi^0$ was found with a very large branching ratio experimentally, which provide evidence for a failure of the naive factorization   and for sizeable relative strong-interaction phases between different isospin amplitudes \cite{neubert}. This was confirmed in the perturbative QCD (PQCD) approach  based on $k_T$ factorization \cite{Keum:2003js, Li:2008ts, Zou:2009zza}, where the endpoint singularity was killed by keeping the transverse momentum of partons.  The rescattering effects of $B\to D^{(*)} M$ had also been studied within some models  \cite{Chua:2007qw}. Under the assumption of the flavor SU(3) symmetry, the global fits were performed in the topological quark diagram approach \cite{Chiang:2007bd}, where the magnitudes and the strong phases of the topologically distinct amplitudes were studied, but the information of SU(3)  asymmetry was lost. Due to the large difference between pseudoscalar and vector meson, their $\chi^2$ fit has to be performed for each category of decays to result in three sets of parameters.

Recently, in order to study the two-body hadronic decays of $D$ mesons, the factorization-assisted topological-amplitude (FAT) approach was proposed \cite{Li:2012cfa, Li:2013xsa}, which combines the conventional factorization approach and topological-amplitude parameterization. We will introduce the the framework in the next section in detail. By involving the non-factorizable contributions and the SU(3) symmetry breaking effect, most theoretical predictions of the $D$ decays are in better agreement with experimental data, and the long-standing puzzle from the $D^0\to\pi^+\pi^-$ and $D^ 0\to K^ + K^-$  branching fractions can be well solved \cite{Li:2012cfa}. In this work, we shall generalize the FAT approach to study the two-body charmed nonleptonic $B$ mesons decays. With the available experimental data for 31 decay channels, we shall fit the only 4 theoretical parameters, reducing from the 15 parameters introduced in \cite{Chiang:2007bd}. The SU(3) asymmetries and their implications will also be discussed. The predicted results  for all the 120 decay channels can be tested in the running LHCb experiment, future Belle-II experiment and even high energy colliders in the future.

This manuscript is organized as follows. In Sec.~\ref{sec:2}, we introduce the framework of FAT approach, and fit the four universal parameters from the available data induced by $b\to c$ transition. In Sec.~\ref{sec:3}, we predict the branching fractions of decays induced by $b\to u$ transition with the assumption that the numerical values of four universal parameters are the same as those of decays of $b\to c$ transition. The discussions on the phenomenological implications will be given in Sec.~\ref{sec:4}. At last, we shall summarize this work in Sec.~\ref{sec:5}.

\section{The CKM favored decays induced by $b\to c$ Transition}\label{sec:2}
\subsection{Framework of FAT Approach}
When discussing the charmed $B$ decays, a new intermediate scale ($m_c$) is introduced, which satisfies the mass hierarchy $m_b > m_c> \Lambda_{QCD}$. The perturbative theory may not be valid in the scale ($m_c$), implying the failure of QCD factorization. Thus the best way is to extract the information of them from experimental data. In the conventional topological diagrammatic approach, the amplitude of each diagram was proposed to be extracted directly \cite{Chiang:2007bd} from experimental data. To achieve this goal, the flavor SU(3) symmetry has to be employed, which works well in the two-body charmless $B$ decays \cite{Chau:1990ay} due to the negligible mass of the light meson. However, in dealing with the $D$ meson decays \cite{Cheng:2010ry}, it is found that only the experimental data of Cabibbo-favored decay modes can be used, which implies that the SU(3) breaking effects are sizable in the $D$ decays. As for the charmed $B$ decays, the effects from SU(3) asymmetry are also expected to be sizable that may not be negligible. Even if people ignore the SU(3) breaking effect of $\pi-K$ difference, the $\chi2$ fit can only be done separately in three categories of decays, namely, $B \to D P$, $B \to D V$, and $B \to D^* P$, with 5 free parameters in each group \cite{Chiang:2007bd}. Obviously, the predictive power is lost with 15 parameters to be fitted from experimental data. With some SU(3) breaking effects input by hand, the number of free parameters  becomes 21 in the $\chi^2$ fit of ref. \cite{Chiang:2007bd}, which is surely not satisfactory.

The factorization-assisted topological-amplitude (FAT) approach was first proposed for studying the two-body hadronic $D$ mesons decays \cite{Li:2012cfa, Li:2013xsa}, which is a great success in the extraction of strong phases for the CP asymmetry study. There are five steps in the FAT approach. Firstly, similar to the topological diagrammatic approach \cite{Chiang:2007bd}, the two-body hadronic weak decay amplitudes are decomposed in terms of some distinct quark diagrams, according to the weak interactions and flavor flows with all strong interaction effects encoded. In this way, the non-negligible non-factorizable contributions are involved, and hence the results would be more accurate if their values can be extracted from experimental data. In the case of charmed hadronic decays of $B$ mesons, four kinds of relevant quark diagrams are involved, namely, the color-favored tree diagram $T$, the color-suppressed tree diagram $C$, the $W$-exchange annihilation-type diagram $E$, and the $W$-annihilation diagram $A$. Secondly, in order to keep the SU(3) breaking effects in the decay amplitudes, we factorize the decay constants and form factors formally from each topological amplitude.  The topological amplitude is then only universal for all decay channels after factorization of those hadronic parameters. Thirdly, the QCD factorization, the perturbative QCD based on $k_T$ factorization, together with the soft-collinear effective theory have all proved factorization  for the color favored topology diagram \cite{Beneke:1999br, Bauer:2001cu, Keum:2003js}. The $T$ amplitude is then safely expressed by the products of transition form factor, decay constant of the emitted meson and the short-distance dynamics Wilson coefficients, where the latter are related to the four-fermion operators. No free parameter will be introduced in the $T$ diagram calculations. Fourthly, for the remaining color suppressed diagram and W-exchange  diagram (W), their size and phase $\chi^C$, $\phi^C$, $\chi^E$ and $\phi^E$ after factorized the decay constants and form factor, are the only four universal free parameters to be fitted from the abundant experimental data simultaneously. Lastly, with the four fitted universal nonperturbative parameters, we then make predictions for  all the hadronic charmed $B$ decays $B_{u,d,s}\to D^{(*)}P(V)$  and $B_{u,d,s}\to \overline D^{(*)}P(V)$, where $P$ and $V$ denote pseudoscalar and vector mesons, respectively.

\begin{figure}[htb]
\begin{center}
\includegraphics[scale=0.25]{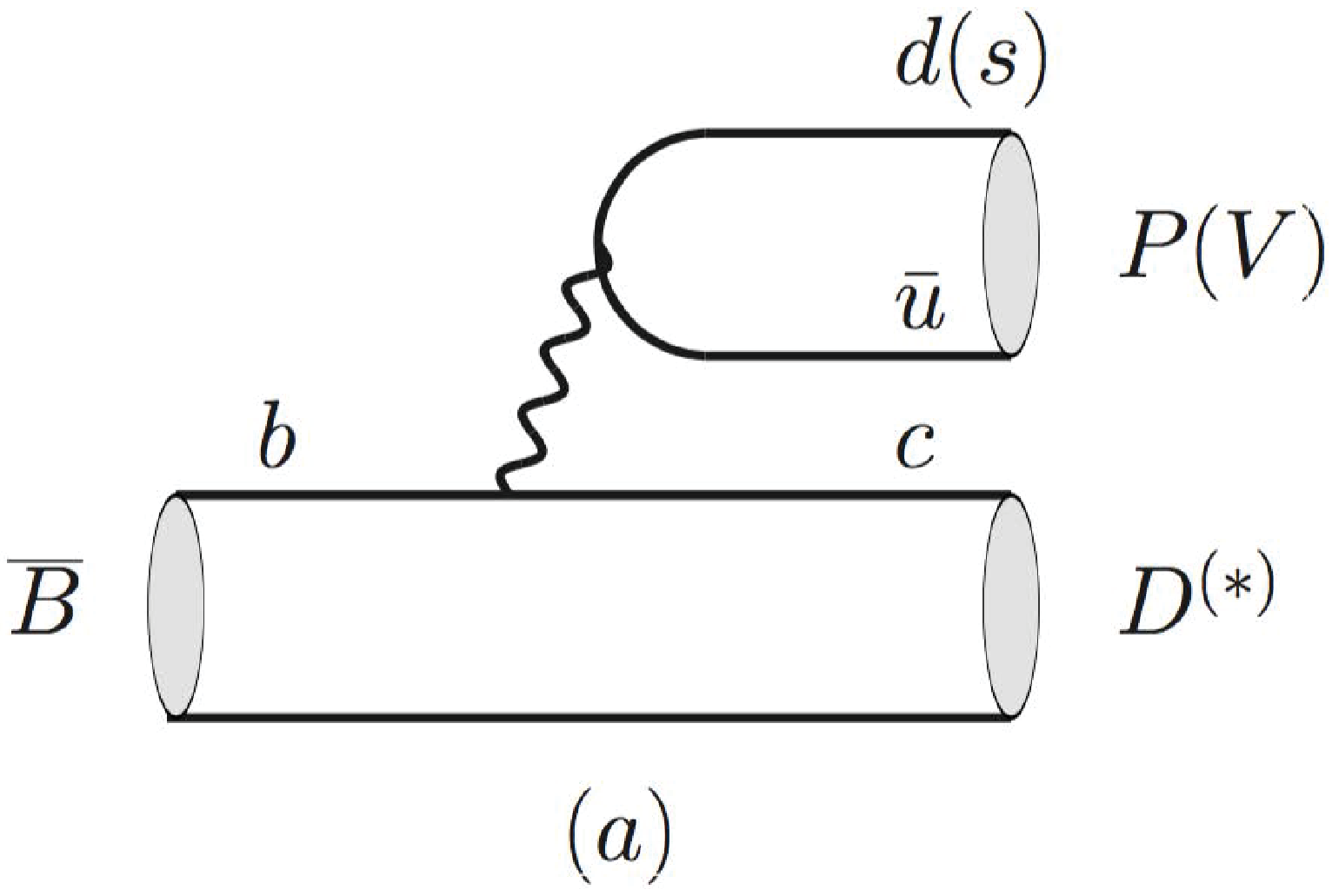}
\includegraphics[scale=0.25]{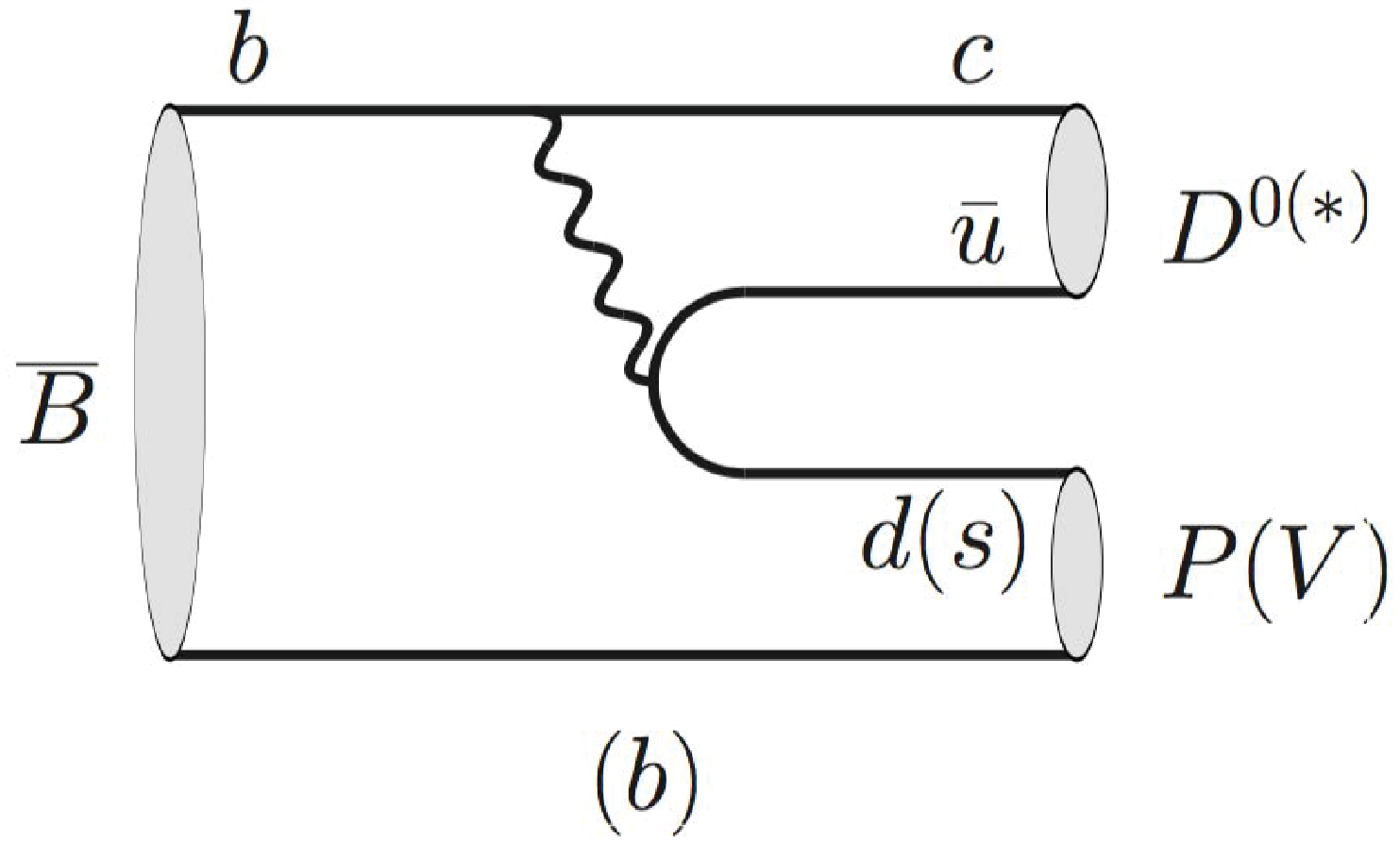}
\includegraphics[scale=0.25]{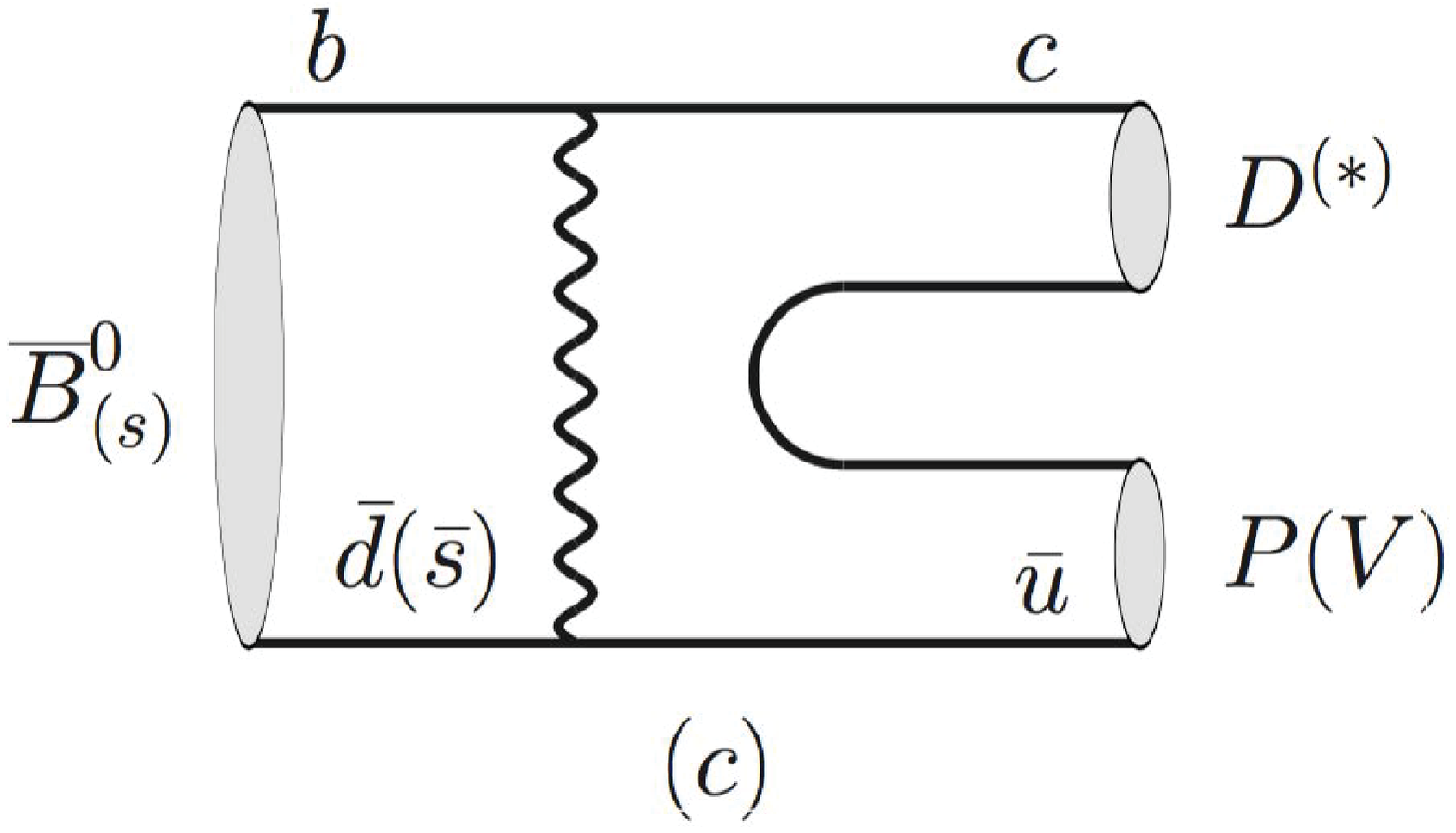}
\end{center}
\caption{Topological diagrams in the $b\to c$ transitions: (a) the color-favored   tree diagram, $T$; (b) the color-suppressed   tree diagram, $C$; and (c) the $W$-exchange annihilation-type diagram, $E$. Note that the $E$ diagram occurs only in the $\overline B_{d}^{0}$ and $\overline B_{s}^{0}$ decays. }\label{fig:top}
\end{figure}

According to the effective Hamiltonian \cite{Buchalla:1995vs}, these decays can be classified into two groups: the CKM favored processes induced by $b\to c$ transition and the CKM suppressed ones induced by $b\to u$ transition.
We firstly discuss the relevant effective weak Hamiltonian for the CKM favored transition $b\to c q \bar u$($q=d,s$), which is given by \cite{Buchalla:1995vs}
\begin{equation}
\mathcal{H}_{\rm eff}  = {G_{F}\over \sqrt 2} V_{cb}V_{uq}^{*}[C_{1}(\mu)O_{1}(\mu)+C_{2}(\mu)O_{2}(\mu)]+h.c.,\label{eq2}
\end{equation}
where $G_{F}$ is the Fermi coupling constant, $V_{cb}$ and $V_{uq}$ are the relevant Cabibbo-Kobayashi-Maskawa (CKM) matrix elements, and $C_{1,2}$ are the Wilson coefficients. The tree-level current-current operators are
\begin{align}
O_{1}=\bar q_{\alpha} \gamma^{\mu}(1-\gamma_{5}) u_{\beta} \bar c_{\beta}\gamma_{\mu}(1-\gamma_{5}) b_{\alpha},
O_{2}=\bar q_{\alpha} \gamma^{\mu}(1-\gamma_{5}) u_{\alpha} \bar c_{\beta}\gamma_{\mu}(1-\gamma_{5}) b_{\beta},
\end{align}
where  $\alpha$ and $\beta$ are the color indices. The topological diagrams in the $b\to c $ transitions includes color-favored tree emission diagram $T$, color-suppressed tree emission $C$, and $W$-exchange diagram $E$, as shown in Fig.\ref{fig:top}. Note that the $W$-annihilation diagram does not occur in the $b\to c$ transition processes, and the $E$ diagram occurs only in the $\overline B_{d}^{0}$ and $\overline B_{s}^{0}$ decays. It is apparent that the $T$ diagram  emits a light meson and recoils a charmed meson, while for the $C$ diagram the charmed meson is emitted and the light meson is recoiled.

In terms of the factorization hypothesis, the three diagrams of the $\overline B \to D P$ modes can be written as
\begin{align}
T_c^{DP}&=i{G_{F}\over\sqrt2}V_{cb}V_{uq}^{*}a_{1}(\mu)f_{P}
(m_{B}^{2}-m_{D}^{2})F_{0}^{B\to D}(m_{P}^{2}),\\
C_c^{DP}&=i{G_{F}\over\sqrt2}V_{cb}V_{uq}^{*} f_{D}(m_{B}^{2}-m_{P}^{2})F_{0}^{B\to P}(m_{D}^{2})\chi_c^{C}e^{i\phi_c^{C}},\\
E_c^{DP}&=i{G_{F}\over\sqrt2}V_{cb}V_{uq}^{*}m_{B}^{2} f_{B}\frac{f_{D_{(s)}}f_{P}}{f_{D}f_{\pi}} \chi_c^{E}e^{i\phi_c^{E}},
\end{align}
where the subscript $c$ stands for the processes induced by $b\to c$ transition, and $f_{P}$ and $f_{D}$ for the decay constants of the pseudoscalar meson and $D$ meson, respectively.  $F_{0}^{B\to D}$ and $F_{0}^{B\to P}$ are the scalar form factors of the  $\overline B\to D$ and $\overline B\to P$ transitions. Here we have followed the conventional Bauer-Stech-Wirbel definition for form factors $F^{BP}_{0,1}$ and $A^{BV}_{0}$ \cite{Wirbel:1985ji}. The inner effective Wilson coefficient is
\begin{eqnarray}\label{eq:a1a2}
a_{1}(\mu)=C_{2}(\mu)+{C_{1}(\mu)\over 3},
\end{eqnarray}
For the $T$ diagram, the non-factorizable contribution is so small that can be ignored safely. On the contrary, for the $C$ diagram, because the factorizable contribution is quite small, the non-factorizable contribution becomes significant. As it belongs to the nonperturbative contribution, we set it as universal and parameterize it as $\chi_c^{C}e^{i \phi_c^C}$, which will be extracted from the experimental data. In principle, the factorizable  scale $\mu$ should be channel dependent, however we find that both the fitted parameters and the predictions are not sensitive to this scale. So, for simplicity, we set $\mu=m_b/2=2.1 \mathrm{GeV}$. The Wilson coefficients $C_1$ and $C_2$ at this scale are  $-0.287$ and $1.132$, respectively. As for the $W$-exchange $E$ diagram, the hadronic parameter $\chi_c^{E}$ and its relative strong phase $\phi_c^{E}$ are     also non-perturbative to be extracted from data. In practice, the dimensionless parameters $\chi_c^{E}$ and $\phi_c^E$ are defined from the $\overline B\to D\pi$ process, to which those for other final states are related via the ratio of the decay constants $(f_{D}f_{P}) /(f_{D^{0}} f_{\pi})$. Obviously, the SU(3) asymmetry also remains in the $E$ diagram. In fact, although the helicity suppression doesn't work with a heavy charm quark in the final state, the factorizable contribution in the $E$ diagram is also negligible due to the smallness of the corresponding Wilson coefficient.

Similarly to the amplitudes of $\overline B\to DP$ decays, the topological amplitudes of $T$, $C$ and $E$ of the $\overline B\to D^{*}P$ and $\overline B\to DV$ decays can be given respectively by
\begin{align}
T_c^{D^{*}P}&=\sqrt2{G_{F}}V_{cb}V_{uq}^{*}a_{1}(\mu)f_{P}m_{D^{*}}A_{0}^{B\to D^{*}}(m_{P}^{2})(\varepsilon^{*}_{D^{*}}\cdot p_{B}),
\\
C_c^{D^{*}P}&=\sqrt2{G_{F}}V_{cb}V_{uq}^{*} f_{D^{*}}m_{D^{*}}F_{1}^{B\to P}(m_{D^{*}}^{2})(\varepsilon^{*}_{D^{*}}\cdot p_{B})\chi_c^{C}e^{i\phi_c^{C}},\\
E_c^{D^{*}P}&=\sqrt2{G_{F}}V_{cb}V_{uq}^{*} m_{D^{*}}f_{B}\frac{ f_{D^{*}_{(s)}}f_{P}}{f_{D}f_{\pi}}\chi_c^{E}e^{i\phi_c^{E}}  (\varepsilon^{*}_{D^{*}}\cdot p_{B});
\end{align}
and
\begin{align}
T_c^{DV}&=\sqrt2{G_{F}}V_{cb}V_{uq}^{*}a_{1}(\mu)f_{V}m_{V}F_{1}^{B\to D}(m_{V}^{2})(\varepsilon^{*}_{V}\cdot p_{B}),
\\
C_c^{DV}&=\sqrt2{G_{F}}V_{cb}V_{uq}^{*} f_{D}m_{V}A_{0}^{B\to V}(m_{D}^{2})(\varepsilon^{*}_{V}\cdot p_{B})\chi_c^{C}e^{i\phi_c^{C}} ,
\\
E_c^{DV}&=\sqrt2{G_{F}}V_{cb}V_{uq}^{*} m_{V} f_{B}\frac{f_{D_{(s)}}f_{V} }{f_{D}f_{\pi}}\chi_c^{E}e^{i\phi_c^{E}}(\varepsilon^{*}_{V}\cdot p_{B}).
\end{align}
In above functions, $\varepsilon^{*}_{D^{*}}$ and $\varepsilon^{*}_{V}$ represent the polarization vectors of the  $D^{*}$ and $V$, and $f_{D^{*}}$ and $f_V$ are the decay constants of the corresponding vector mesons. $F_{1}^{B\to D}$ and $F_{1}^{B\to P}$ stand for the vector form factors of $\overline B\to D$ and $\overline B\to P$ transitions, $A_{0}^{B\to D^{*}}$ and $A_{0}^{B\to V}$ are the transition form factors of $B\to D^{*}$ and $\overline B\to V$. Note that, after factorizing the corresponding form factors and decay constants, we can use the same non-perturbative universal parameters for all the $\overline B\to DP$,   $\overline B\to D^{*}P$ and $\overline B\to DV$ decays. The total number of free parameters to be fitted from experimental data remains four. This is contrast to the conventional topological diagram approach \cite{Chiang:2007bd}, where 15 parameters needed for the three categories of processes.

In a short summary, utilizing the factorization, the color favored tree diagram, which is the dominant contribution in many decay channels, is determined by perturbative calculations. For the color suppressed tree diagram and W-exchange diagram, we have only four universal non-perturbative parameters, namely $\chi_{c}^{C}$, $\phi_c^{C}$, $\chi_c^{E}$, and $\phi_c^{E}$ to be fitted from all available $\overline B\to D P$, $D^{*} P$ and $D V$ modes. As stated, most SU(3) breaking effects are involved in the decay constants and the transition form factors. Using the parameters determined from data, we can also reproduce branching fractions of $\overline B\to D P$, $D^{*} P$ and $D V$ modes.
\subsection{Input Parameters}
In this section, we list the used parameters, such as CKM matrix elements, decay constants and transition form factors. Since all the decay modes discussed are induced by the tree level electroweak diagrams, we need not the weak phases of the CKM matrix elements, but use their averaged values of the magnitudes in PDG \cite{PDG}:
\begin{align}
&|V_{cb}|=0.041,~~~|V_{us}|=0.225,~~~|V_{ud}|=0.974,\\
&|V_{ub}|=0.00413 ,~~~|V_{cs}|=0.986 ,~~~|V_{cd}|=0.225.
\end{align}
The decay constants of $\pi$, $K$, $D$ and $D_{s}$ are given by PDG \cite{PDG}. The decay constants of other mesons can not be obtained from experiments directly but calculated in several theoretical approaches, such as the quark model \cite{Verma:2011yw}, the covariant light front approach \cite{Cheng:2003sm}, the light-cone sum rules \cite{Ball:2006eu,Straub:2015ica}, the QCD sum rules \cite{Jamin:2001fw, Gelhausen:2013wia,Lucha:2014xla,Penin:2001ux,Narison:2001pu, Lucha:2010ea,Narison:2012xy}, and the lattice QCD \cite{Dowdall:2013tga,Carrasco:2013naa,Dimopoulos:2011gx, McNeile:2011ng,Bazavov:2011aa,Na:2012kp, Bussone:2014bya,Bernardoni:2014fva} etc. Since the numerical values are different in   different theoretical approaches, we choose the values shown in Table.~\ref{tab:dc} and keep a $5\%$ uncertainty of them.

\begin{table}[tbhp]
\caption{The decay constants of mesons (in unit of MeV). }
\begin{center}
\begin{tabular}{ccccccccccccccc}
\hline\hline
$f_{B}$&
$f_{B_{s}}$&
$f_{D}$&
$f_{D_{s}}$&
$f_{D^{*}}$&
$f_{D_{s}^{*}}$&
$f_{\pi}$&
$f_{K}$&
$f_{\rho}$&
$f_{K^{*}}$&
$f_{\omega}$&
\\\hline
190&
225&
205&
258&
220&
270&
130&
156&
215&
220&
190&
\\
\hline\hline
\end{tabular}\label{tab:dc}
\end{center}
\end{table}

Due to the absence of enough experimental data, the transition form factors of $B$ meson decays have been calculated in the theoretical approaches, such as constitute quark model and light cone quark model \cite{Verma:2011yw, Melikhov:2000yu, Geng:2001de, Lu:2007sg, Albertus:2014bfa}, covariant light front approach(LFQM) \cite{Cheng:2003sm, Cheng:2009ms, Chen:2009qk}, light-cone sum rules \cite{Straub:2015ica,Ball:1998kk, Ball:1998tj, Ball:2001fp,  Ball:2004ye, Ball:2004rg, Ball:2007hb, Charles:1998dr, Beneke:2000wa, Bharucha:2010im, Bharucha:2012wy, Khodjamirian:2006st, Khodjamirian:2011ub,Wang:2015vgv, Meissner:2013hya, Wu:2006rd, Wu:2009kq, Duplancic:2008ix, Ivanov:2011aa, Ahmady:2014sva, Fu:2014uea}, PQCD \cite{Li:2012nk,Wang:2012ab, Wang:2013ix, Fan:2013qz, Fan:2013kqa, Kurimoto:2001zj, Kurimoto:2002sb,Lu:2002ny, Wei:2002iu, Huang:2004hw}, and lattice QCD \cite{Horgan:2013hoa,Dalgic:2006dt, Lellouch:2011qw,Aoki:2013ldr} etc. Considering all above results, we list the the maximum-recoil form factors in Table. \ref{tab:ff}. When dealing with the nonleptonic $B$ decays, we indeed need the form factors with $q^2$ dependence. In order to describe the $q^2$-dependence of form factors, several types of parametrization are proposed. In the current work, we use the dipole parametrization:
\begin{equation}\label{eq:ffdipole}
F_{i}(q^{2})={F_{i}(0)\over 1-\alpha_{1}{q^{2}\over m_{\rm pole}^{2}}+\alpha_{2}{q^{4}\over m_{\rm pole}^{4}}},
\end{equation}
where $F_{i}$ denotes $F_{0}$, $F_{1}$, and $A_{0}$, and $m_{\rm pole}$ is the mass of the corresponding pole state, such as $B$ for $A_{0}$, and $B^{*}$ for $F_{0,1}$. The values of $\alpha_1$ and $\alpha_2$ are also given in Table \ref{tab:ff}. In fact, numerical results show that the $q^2$ dependence of form factors makes little change in our numerical calculations.

For the decay modes with $\eta$ or $\eta'$ in the final state, it is convenient to consider the flavor mixing of $\eta_{q}$ and $\eta_{s}$, defined by
\begin{align}
\eta_{q}={1\over\sqrt2}(u\bar u+ d\bar d),~~~~~\eta_{s}=s\bar s.
\end{align}
Then, $\eta$ and $\eta'$ are linear combinations of $\eta_{q}$ and $\eta_{s}$,
\begin{equation}
\begin{pmatrix} \eta \\ \eta' \end{pmatrix}
=\begin{pmatrix} \cos\phi & -\sin\phi \\ \sin\phi & \cos\phi \end{pmatrix}
\begin{pmatrix} \eta_{q} \\ \eta_{s} \end{pmatrix} ,
\end{equation}
where the mixing angle is determined to be $\phi=(40.4\pm0.6)^{\circ}$ by KLOE \cite{Ambrosino:2009sc}. The flavor specific decay constants  are $f_{q}=(1.07\pm0.02)f_{\pi}$ and $f_{s}=(1.34\pm0.06)f_{\pi}$, corresponding to $\eta_{q}$ and $\eta_{s}$ respectively \cite{Feldmann:1998vh,Feldmann:1998sh}. In this work, the small effect from the mixing between $\omega$ and $\phi$ is ignored.

Honestly, some form factors and decay constants occur only in special channels, so their numerical values would affect the accuracy of our theoretical predictions. In this article, in order to estimate the uncertainties maximally, we shall assign the uncertainties of form factors to be $10\%$, and the uncertainties of decay constants to be $5\%$. If we can determine the form factors and the decay constants more precisely by the experimental data in the future, the predicted results in the FAT approach would be improved.

\begin{table}[hbt]
\caption{The transition form factors at maximum recoil and dipole model parameters used in this work. }\label{tab:ff}
\centering
\begin{tabular}{|c||c|c|c|c|c||c|c|c|c|c|c|c|}
\hline&
$F_{0}^{B\to\pi}$&
$F_{0}^{B\to K}$ &
$F_{0}^{B_{s}\to K}$&
$F_{0}^{B\to\eta_{q}}$&
$F_{0}^{B_{s}\to\eta_{s}}$&
$F_{0}^{B\to D}$&
$F_{0}^{B_{s}\to D_{s}}$&
$A_{0}^{B\to D^{*}}$&
$A_{0}^{B_{s}\to D^{*}_{s}}$&
$F_{1}^{B \to D}$&
$F_{1}^{B_{s} \to D_{s}}$ \\
\hline
$F(0)$&
0.28&
0.33&
0.29&
0.21&
0.31&
0.54&
0.58&
0.56&
0.57&
0.54&
0.58\\
$\alpha_1$&
0.50&
0.53&
0.54&
0.52&
0.53&
1.71&
1.69&
2.44&
2.49&
2.44&
2.44\\
$\alpha_2$&
-0.13&
-0.13&
-0.15&
0    &
0    &
0.52 &
0.78 &
1.98 &
1.74 &
1.49 &
1.70\\
\hline
\hline&
$F_{1}^{B\to\pi}$&
$F_{1}^{B\to K}$&
$F_{1}^{B_{s}\to K}$&
$F_{1}^{B \to\eta_{q}}$&
$F_{1}^{B_{s}\to\eta_{s}}$&
$B_{(s)}\to V$&
$A_{0}^{B \to\rho}$&
$A_{0}^{B\to \omega}$&
$A_{0}^{B_{s}\to \phi}$&
$A_{0}^{B\to K^{*}}$&
$A_{0}^{B_{s}\to K^{*}}$\\
\hline
$F(0)$&
0.28&
0.33&
0.29&
0.21&
0.31&
$A(0)$&
0.30&
0.26&
0.30&
0.33&
0.27\\
$\alpha_1$ &
0.52&
0.54&
0.57&
1.43&
1.48&
$\alpha_1$ &
1.56&
1.60&
1.73&
1.51&
1.74\\
$\alpha_2$ &
0.45&
0.50&
0.50&
0.41&
0.46&
$\alpha_2$&
0.17&
0.22&
0.41&
0.14&
0.47\\
\hline
\hline
\end{tabular}
\end{table}

\subsection{$\chi^{2}$ Fit}
As discussed above, there are only four parameters in the FAT approach, namely $\chi^{C}$, $\phi^{C}$, $\chi^{E}$ and $\phi^{E}$, which are universal to all $\overline B\to DP$, $D^{*}P$ and $DV$ decays. In the fitting, we define the $\chi^2$ function in term of $n$ experimental observables $x_i\pm\Delta x_i$ and the corresponding theoretical predictions $x_i^{\rm th}$,
\begin{align}
\chi^{2}=\sum_{i=1}^{n}\left(\frac{x_i^{\rm th}-x_i}{\Delta x_i}\right)^2 .
\end{align}
In this work, the data points are the branching fractions. We then write the corresponding theoretical predictions in terms of topological amplitudes and extract the four parameters by minimizing $\chi^2$. Currently, there are 31 experimental measured charmed decay modes induced by $b\to c$ transition \cite{PDG}. With these data, the best-fitted values of the parameters are obtained as
\begin{align}
\chi_c^{C}=0.48\pm0.01,~~~
\phi_c^{C}=(56.6^{+3.2}_{-3.8})^{\circ},~~~
\chi_c^{E}=0.024^{+0.002}_{-0.001},~~~
\phi_c^{E}=(123.9^{+3.3}_{-2.2})^{\circ},
\end{align}
with $\chi^{2}/d.o.f.=1.4$.

In ref.~\cite{Chiang:2007bd}, Chiang $et.al$ fitted the amplitudes and strong phases of each diagrams using the latest experimental data in the topological diagram approach. Because they do not include the SU(3) breaking effects properly, they had to fit each amplitude of $B \to DP, DV$ and $D^*P$ decays separately. Even though with much more parameters than us, their $\chi^2$ per degree of freedom is larger than ours. Only under the so-called scheme 3, where some of the SU(3) symmetry breaking effects have been involved, 21 parameters to be fitted from data, their $\chi^2/d.o.f.$ for the $B \to DP$ decays is a little smaller than ours. But the $\chi^2/d.o.f.$ for the $B \to DV$ and $D^*P$ decays is still larger than ours. With so many parameters, they lost the predictive power of the branching fractions, because there are not enough data of $B \to \overline D^{(*)}M$ decays. By contrast, we can predict 120 branching fractions, by fitting 4 parameters from 31 decay modes.

\subsection{Branching Fractions}
With the fitted parameters, the topological amplitudes and the predicted branching fractions of $\overline B\to DP$, $D^{*}P$ and $DV$ decays induced by $b\to c$ transition are shown in Tables~\ref{tab:BrbcDP}, \ref{tab:BrbcDstP} and \ref{tab:BrbcDV}, respectively. The experimental data are also given for comparison. For all theoretical predictions, the first uncertainties arise from the  aforementioned four parameters fitted in the FAT approach. The second uncertainties come from the transition form factors, and the third ones are from decay constants. From the tables, it is obvious that the major uncertainties are from form factors Moreover, we note that each table is divided into two parts, Cabibbo-favored ($V_{ud}$ or $V_{cs}$) and Cabibbo-suppressed ($V_{us}$ or $V_{cd}$), and most branching fractions of the Cabibbo-favored processes are larger than those of Cabibbo-suppressed ones.

From the tables, we find that our results are consistent with the measured $B^- $ and $\overline B^0 $ decays induced by $b\to c$ transition. As for $\overline B_s^0$,  only a few typical decays, such as $\overline B_s^0 \to D_s^{(*)+}\pi^-$, have been measured in LHCb, while most of them will be tested in LHCb in the following years. Comparing with ref.\cite{Chiang:2007bd}, most of the results are in agreement with each other.

\begin{table}[tbh!]
\caption{Branching fractions and decay amplitudes for the $\overline B\to DP$ modes. Data are from \cite{PDG}. The first uncertainties are from the fiited parameters, the second uncertainties are from the form factors, and the third ones are coming from decay constants. }\label{tab:BrbcDP}
\begin{center}
\begin{tabular}{lllllll}
\hline\hline
Meson~~~~~~~~ &
Mode~~~~~~~~ &
Amplitudes ~~~~~~~~ &
$\mathcal{B}_{\rm exp}(\times10^{-4})$~~~~~~~~  &
$\mathcal{B}_{\rm th}(\times10^{-4})$~~~~~~~~\\
\hline &
Cabibbo-favored &
$V_{cb}V_{ud}^{*}$\\
\hline$
\overline B^{0}$&
$D^{+}\pi^{-}$&
$T+E$&
$26.8\pm1.3$&
$24.7^{+0.2}_{-0.1}\pm5.1\pm0.1$
\\&
$D^{0}\pi^{0}$&
${1\over\sqrt2}(E-C)$&
$2.6\pm0.1$&
$2.5^{+0.1}_{-0.2}\pm0.5\pm0.1$
\\&
$D^{0}\eta$&
${1\over\sqrt2}(C+E)\cos\phi$&
$2.4\pm0.3$&
$1.9\pm0.1\pm0.4\pm0.1$
\\&
$D^{0}\eta'$&
${1\over\sqrt2}(C+E)\sin\phi$ &
$1.38\pm0.16$&
$1.3\pm 0.1\pm0.2\pm 0.1$
\\&
$D_{s}^{+}K^{-}$&
$E$&
$0.345\pm$0.032 &
$0.30^{+0.04}_{-0.02}\pm0.00\pm0.03$
\\
\hline
$B^{-}$&
$D^{0}\pi^{-}$&
$T+C$&
$48.1\pm1.5$&
$49.0^{+1.4}_{-1.7}\pm7.6\pm 0.6$
\\
\hline
$\overline B_{s}^{0}$&
$D_{s}^{+}\pi^{-}$&
$T$&
$30.4\pm2.3$&
$30.2\pm{0.0}\pm6.0\pm0.1$
\\&
$D^{0}K^{0}$&
$C$&  &
$5.9\pm0.3\pm 1.2\pm 0.3$\\
\hline
\hline
& Cabibbo-suppressed &
$V_{cb}V_{us}^{*}$\\\hline
$\overline B^{0}$ &
$D^{+}K^{-}$ &
$T$ &
$1.97\pm0.21$ &
$2.1\pm 0.0\pm0.4\pm 0.0$\\&
$D^{0}\overline K^{0}$ &
$C$ &
$0.5\pm0.1$ &
$0.4\pm0.0\pm0.1\pm0.0$\\\hline
$B^{-}$&
$D^{0}K^{-}$&
$T+C$&
$3.70\pm0.17$&
$3.8\pm0.1\pm 0.6\pm 0.1$\\ \hline
$\overline B_{s}^{0}$&
$D_{s}^{+}K^{-}$&
$T+E$&  &
$2.1\pm0.0\pm0.4\pm0.0$\\&
$D^{0}\eta$ &
${1\over\sqrt2}E\cos\phi-C\sin\phi$ &  &
$0.14\pm 0.01\pm 0.03\pm 0.01$\\&
$D^{0}\eta'$ &
${1\over\sqrt2}E\sin\phi+C\cos\phi$ &  &
$0.21\pm 0.01\pm 0.04\pm 0.01$\\&
$D^{+}\pi^{-}$&
$E$ &  &
$0.011\pm0.001\pm 0.000\pm0.001$\\&
$D^{0}\pi^{0}$ &
${1\over\sqrt2}E$ &  &
$0.005^{+0.001}_{-0.000}\pm 0.000\pm 0.001$\\
\hline\hline
\end{tabular}
\end{center}
\end{table}

\begin{table}[tbph]
\caption{Branching fractions and decay amplitudes for the $\overline B\to D^{*}P$ decays.}\label{tab:BrbcDstP}
\begin{center}
\begin{tabular}{lllllll}
\hline\hline
Meson~~~~~~~~ & Mode~~~~~~~~ &Amplitudes ~~~~~~~~ &$\mathcal{B}_{\rm exp}(\times10^{-4})$~~~~~~~~  &$\mathcal{B}_{\rm th}(\times10^{-4})$~~~~~~~~\\
\hline&
Cabibbo-favored &
$V_{cb}V_{ud}^{*}$\\
\hline
$\overline B^{0}$ &
$D^{*+}\pi^{-}$ &
$T+E$ &
$27.6\pm1.3$ &
$24.9^{+0.2}_{-0.1}\pm 5.2\pm 0.1$\\ &
$D^{*0}\pi^{0}$ &
${1\over\sqrt2}(E-C)$ &
$2.2\pm0.6$ &
$2.8\pm{ 0.2} \pm 0.6\pm 0.3$\\ &
$D^{*0}\eta$ &
${1\over\sqrt2}(C+E)\cos\phi$&
$2.3\pm0.6$ &
$2.1\pm 0.1\pm 0.4\pm 0.2$\\ &
$D^{*0}\eta'$ &
${1\over\sqrt2}(C+E)\sin\phi$&
$1.40\pm0.22$ &
$1.4\pm0.1\pm 0.2\pm 0.1$\\ &
$D_{s}^{*+}K^{-}$ &
$E$ & $0.219\pm0.030$ &
$0.22^{+ 0.03}_{- 0.01}\pm 0.00\pm 0.03$\\
\hline
$B^{-}$ &
$D^{*0}\pi^{-}$ &
$T+C$ & $51.8\pm2.6$  &
$50.7^{+ 1.5}_{-1.8} \pm 7.8\pm 1.4$\\
\hline
$\overline B_{s}^{0}$ &
$D_{s}^{*+}\pi^{-}$ &
$T$ & $20\pm5$  &
$27.1\pm 0.0\pm 5.4\pm 0.1$\\ &
$D^{*0}K^{0}$ &
$C$ &  &
$6.6^{+0.3}_{-0.4} \pm 1.3 \pm 0.7$\\
\hline
\hline& Cabibbo-suppressed &
$V_{cb}V_{us}^{*}$\\
\hline
$\overline B^{0}$ &
$D^{*+}K^{-}$ &
$T$ &
$2.14\pm0.16$ &
$2.0\pm 0.00\pm 0.4\pm 0.0$\\ &
$D^{*0}\overline K^{0}$ &
$C$ &
$0.36\pm0.12$ &
$0.45^{+ 0.02}_{- 0.03}\pm 0.09\pm 0.05$\\
\hline
$B^{-}$ &
$D^{*0}K^{-}$ &
$T+C$ &
$4.20\pm0.34$ &
$3.8\pm0.1\pm 0.6\pm 0.1$\\ \hline
$\overline B_{s}^{0}$ &
$D_{s}^{*+}K^{-}$ &
$T+E$ &  &
$1.9\pm0.0\pm 0.4\pm 0.0$\\&
$D^{*0}\eta$ &
${1\over\sqrt2}E\cos\phi-C\sin\phi$ &  &
$0.15\pm 0.01\pm 0.03\pm 0.02$\\&
$D^{*0}\eta'$ &
${1\over\sqrt2}E\sin\phi+C\cos\phi$ &  &
$0.23\pm 0.01\pm 0.04\pm 0.02$\\ &
$D^{*+}\pi^{-}$ &
$E$ &
$<0.061$ &
$0.008\pm0.001\pm 0.000\pm 0.001$\\&
$D^{*0}\pi^{0}$ &
${1\over\sqrt2}E$ &  &
$0.004^{+ 0.004}_{-0.000}\pm 0.000\pm 0.001$\\
\hline\hline
\end{tabular}
\end{center}
\end{table}
\begin{table}[tbph]
\caption{Branching fractions and decay amplitudes for the $\overline B\to DV$ decays.}\label{tab:BrbcDV}
\begin{center}
\begin{tabular}{lllllcl}
\hline\hline
Meson~~~~~~~~ & Mode~~~~~~~~ &Amplitudes ~~~~~~~~ &$\mathcal{B}_{\rm exp}(\times10^{-4})$~~~~~~~~  &$\mathcal{B}_{\rm th}(\times10^{-4})$~~~~~~~~\\
\hline
& Cabibbo-favored & $V_{cb}V_{ud}^{*}$\\
\hline$\overline B^{0}$ & $D^{+}\rho^{-}$ & $T+E$ & $78\pm13$ &$65.3^{+0.5}_{-0.3}\pm13.5\pm 6.6 $\\
& $D^{0}\rho^{0}$ & ${1\over\sqrt2}(E-C)$ & $3.2\pm0.5$ &$ 2.6\pm 0.2\pm 0.6\pm 0.1 $\\
& $D^{0}\omega$ & ${1\over\sqrt2}(E+C)$ & $2.54\pm0.16$ &$ 2.7\pm0.2\pm 0.5\pm 0.1$\\
& $D_{s}^{+}K^{*-}$ & $E$ & $0.35\pm0.10$ &$ 0.38^{+ 0.05}_{- 0.02}\pm 0.00\pm 0.06$\\
\hline$B^{-}$ & $D^{0}\rho^{-}$ & $T+C$ & $134\pm18$  &$ 105^{+2}_{-3} \pm 18 \pm 9$\\
\hline$\overline B_{s}^{0}$ & $D_{s}^{+}\rho^{-}$ & $T$ & $70\pm15$  &$78.6\pm0.0\pm 15.7\pm7.9$\\
& $D^{0}K^{*0}$ & $C$ & $3.5\pm0.6$  &$ 4.9^{+0.2}_{-0.3} \pm 1.0 \pm 0.2 $\\ \hline
\hline
& Cabibbo-suppressed & $V_{cb}V_{us}^{*}$\\
\hline$\overline B^{0}$ & $D^{+}K^{*-}$ & $T$ & $4.5\pm0.7$ &$ 3.9\pm  0.0 \pm 0.8 \pm 0.4 $\\
& $D^{0}\overline K^{*0}$ & $C$ & $0.42\pm0.06$ &$ 0.37\pm0.02\pm 0.07\pm 0.02$\\
\hline$B^{-}$ & $D^{0}K^{*-}$ & $T+C$ & $5.3\pm0.4$ &$ 6.0^{+0.1}_{-0.2}\pm 1.0 \pm 0.5 $\\
\hline$\overline B_{s}^{0}$ & $D_{s}^{+}K^{*-}$ & $T+E$ &  &$4.0^{+0.04}_{-0.03}\pm 0.8\pm 0.4$\\
& $D^{0}\phi$ & $C$ & $0.24\pm0.07$ &$ 0.31^{+ 0.01}_{- 0.02}\pm 0.06\pm 0.02$\\
& $D^{+}\rho^{-}$ & $E$ &  &$ 0.019^{+ 0.002}_{- 0.001}\pm 0.000\pm 0.003$\\
& $D^{0}\rho^{0}$ & ${1\over\sqrt2}E$ &  &$ 0.010\pm0.001\pm 0.000\pm 0.001$\\
& $D^{0}\omega$   & ${1\over\sqrt2}E$ &  &$ 0.008\pm0.001\pm 0.000\pm 0.001$\\

\hline\hline
\end{tabular}
\end{center}
\end{table}
In Tables~\ref{tab:BrbcDP}, \ref{tab:BrbcDstP} and \ref{tab:BrbcDV}, for the decays dominated by the $T$ diagram, because the decay constants of light vector mesons are much larger than those of light pseudoscalar ones, the branching fractions of the $\overline B\to DV$ decays are larger than those of the $\overline B\to DP$ and $\overline B\to D^{*}P$  with light meson emitted. For example, the branching fraction of $\overline B^0 \to D^+\rho^-$ are larger than those of $\overline B^0 \to D^{(*)+}\pi^-$ by a factor of 2.6 because of $f_\rho > f_\pi$. Similarly, we obtain $\mathcal{B}(\overline B^{0}\to D^{+}K^{*-}) > \mathcal{B}(\overline B^{0}\to D^{(*)+}K^{-})$,  $\mathcal{B}(\overline B_s^{0}\to D_s^{+}\rho^{-}) > \mathcal{B}(\overline B_s^{0}\to D_s^{(*)+}\pi^{-})$  and $\mathcal{B}(\overline B_s^{0}\to D_s^{+}K^{*-}) > \mathcal{B}(\overline B_s^{0}\to D_s^{(*)+}K^{-})$. For the $D^*P$ modes, there is no contribution of transverse polarizations and the behavior of the longitudinal polarization is similar to that of the pseudoscalar meson, so the branching fractions of $\overline B\to D^*P$ are close to those of $\overline B\to DP$.

Compared with the QCD-inspired methods \cite{Beneke:1999br, Bauer:2001cu, Li:2008ts, Zou:2009zza}, the amplitudes of color-suppressed $C$ diagrams are relatively large in the FAT approach where the non-factorizable contribution are dominant, as well as in the topological approach \cite{Chiang:2007bd}. From   Table~\ref{tab:BrbcDP}, it is found that the branching fraction of $\overline B^0 \to D^{+}\pi^-$ is larger than that of $\overline B^0 \to D_s^{+}K^-$ by two orders of magnitude, which implies that the contribution of $E$ diagram is much smaller than that of $T$ diagram. So, the $E$ diagram can be neglected as a good approximation in the processes with both $T$ and $E$ contributions. In the comparison between the $\overline B^0 \to D^{+} K^-$ with $\overline B^0 \to D^0\overline K^0$, and $\overline B_s^0 \to D_s^{+}\pi^-$ with $\overline B_s^0 \to D^{0}K^0$, we find that
\begin{align}
|C_c^{DP}|/|T_c^{DP}|\sim 0.45.
\end{align}
Then, the hierarchy
\begin{align}
|T_c^{DP}|:|C_c^{DP}|:|E_c^{DP}|\sim 1:0.45:0.1
\end{align}
are obtained in the FAT approach. Similarly, we also get
\begin{align}
|T_c^{D^*P}| :|C_c^{D^*P}| :|E_c^{D^*P}| \sim 1:0.36:0.1\\
|T_c^{DV}|:|C_c^{DV}|:|E_c^{DV}|\sim 1:0.31:0.1.
\end{align}
It is obvious that these relations  differ from the relation $|T_c^{DP}|\gg |C_c^{DP}| \sim |E_c^{DP}|$ arrived in the PQCD approach \cite{Li:2008ts}, which have  significant impacts on the processes without $T$ diagrams. For example, the topological amplitudes of  $\overline B^{0}\to D^{0}\rho^{0}$ and $D^{0}\omega$ decays are $(E-C)/\sqrt2$ and $(E+C)/\sqrt2$, respectively. The branching fraction of the $D^{0}\rho^{0}$ mode is predicted to be almost one half of that of the $D^{0}\omega$ mode in the PQCD approach \cite{Li:2008ts}, since $C$ and $E$ diagrams contribute destructively for the former mode but constructively for the latter one, which does not agree with  the experiment. However, this issue can be easily explained in the FAT approach in which both   channels are  dominated by the $C$ diagram. With the same argument, the experimental data of the decay modes $\overline B_d^{0}\to D^{0}\pi^{0}$ and $D^{0}\overline K^{0}$ can be easily understood.

\section{The CKM suppressed decays induced by $b\to u$ transition}\label{sec:3}
\begin{figure}[phtb]
\begin{center}
\includegraphics[scale=0.25]{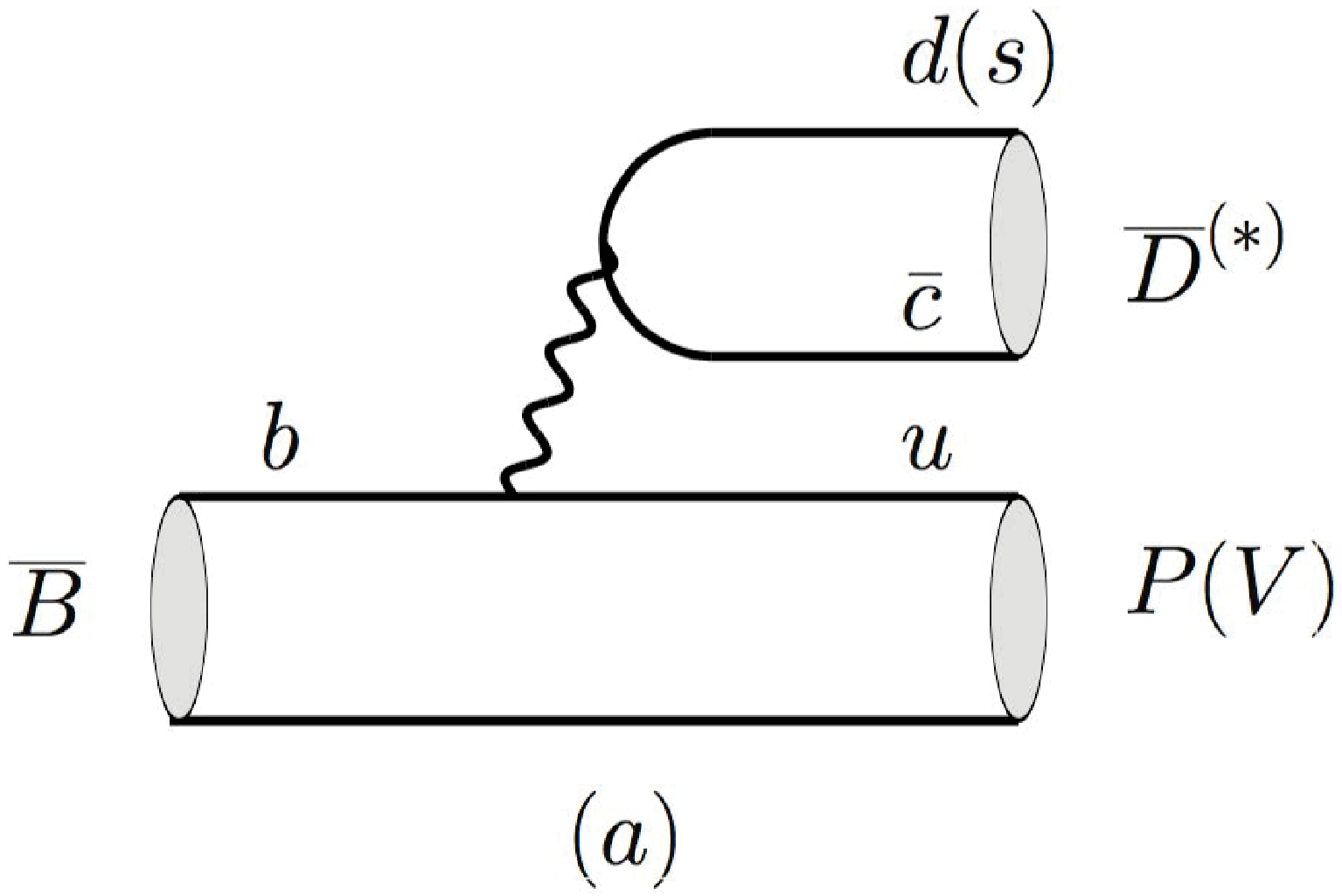}
\includegraphics[scale=0.25]{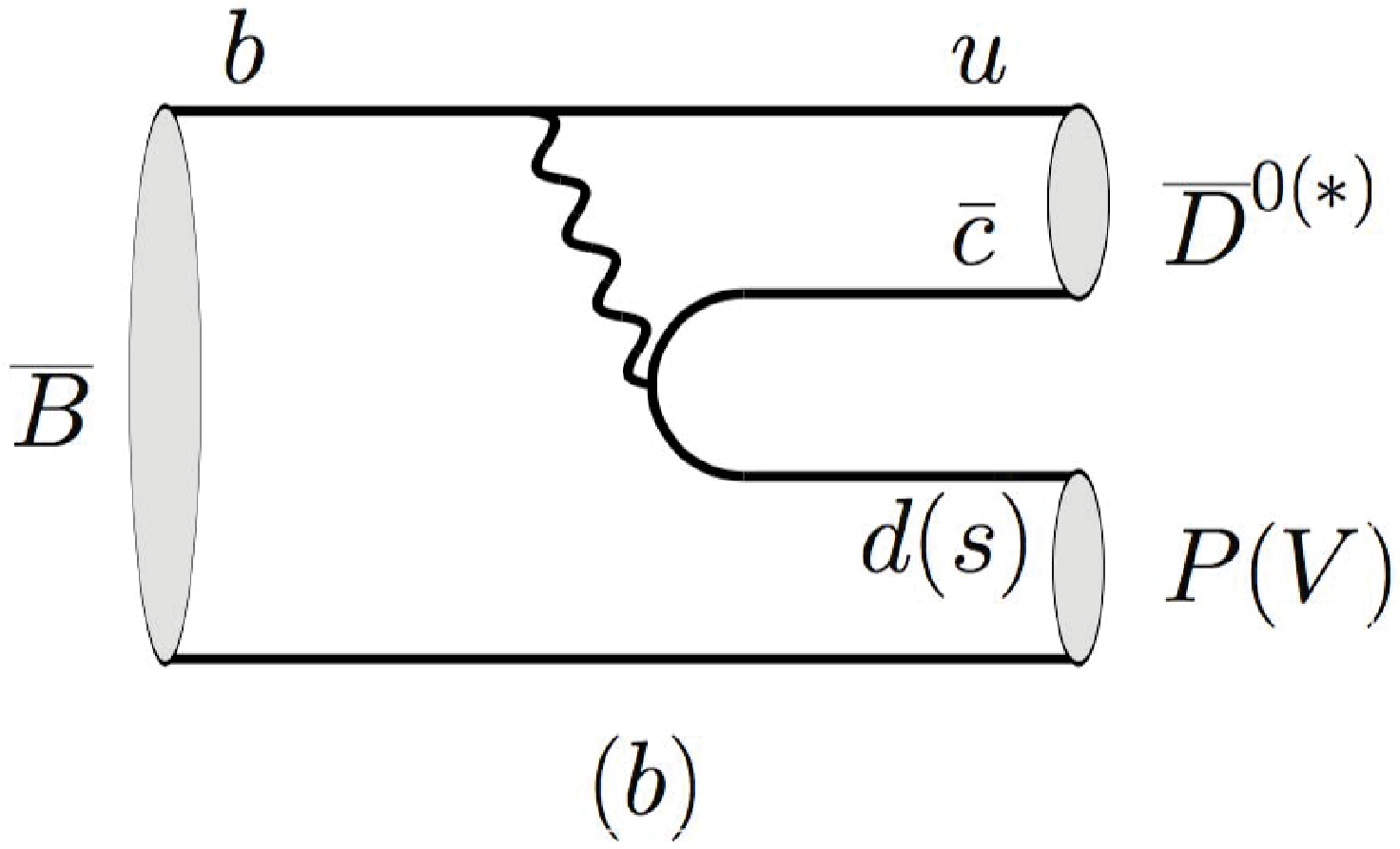}
\includegraphics[scale=0.25]{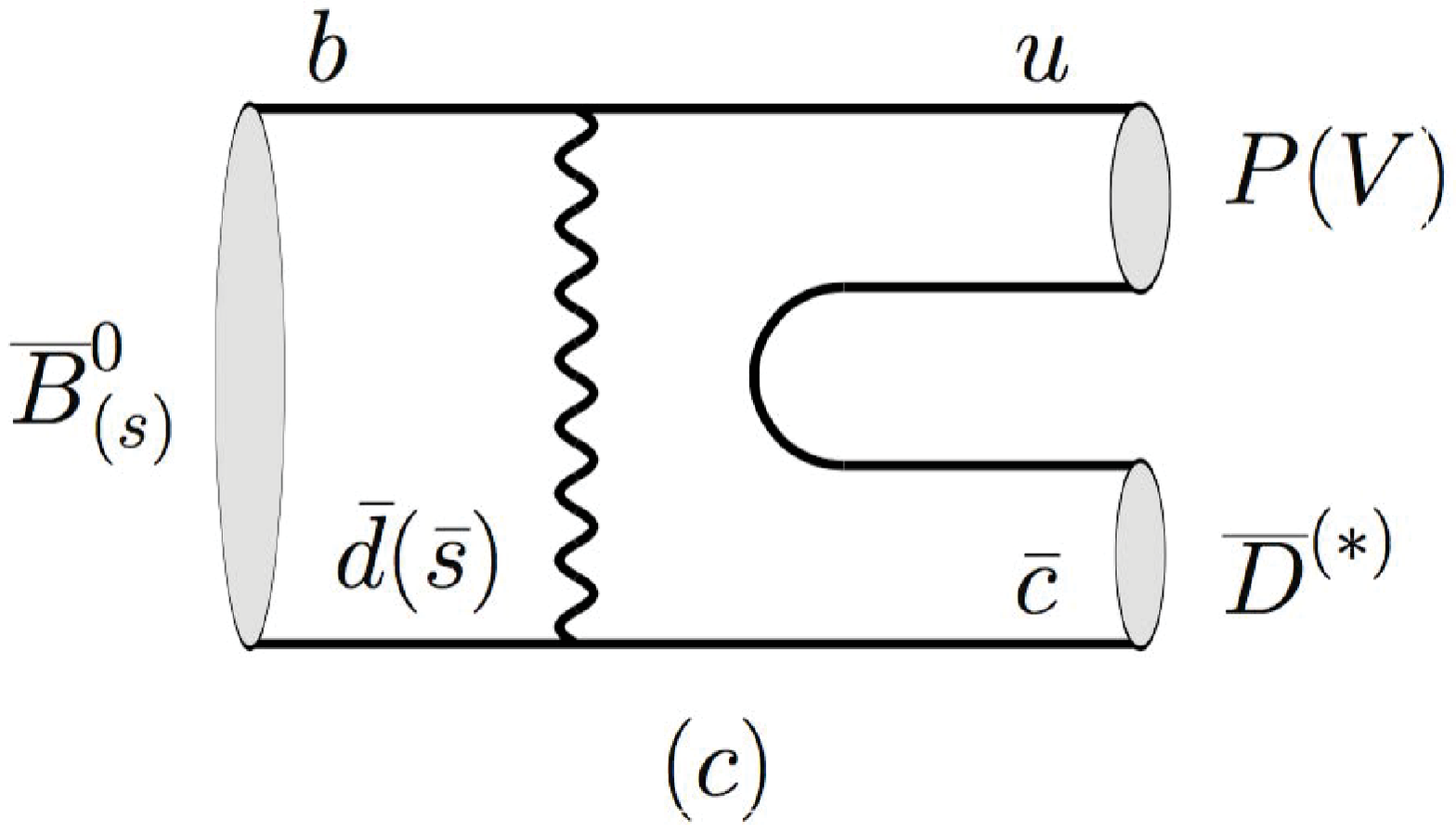}
\includegraphics[scale=0.25]{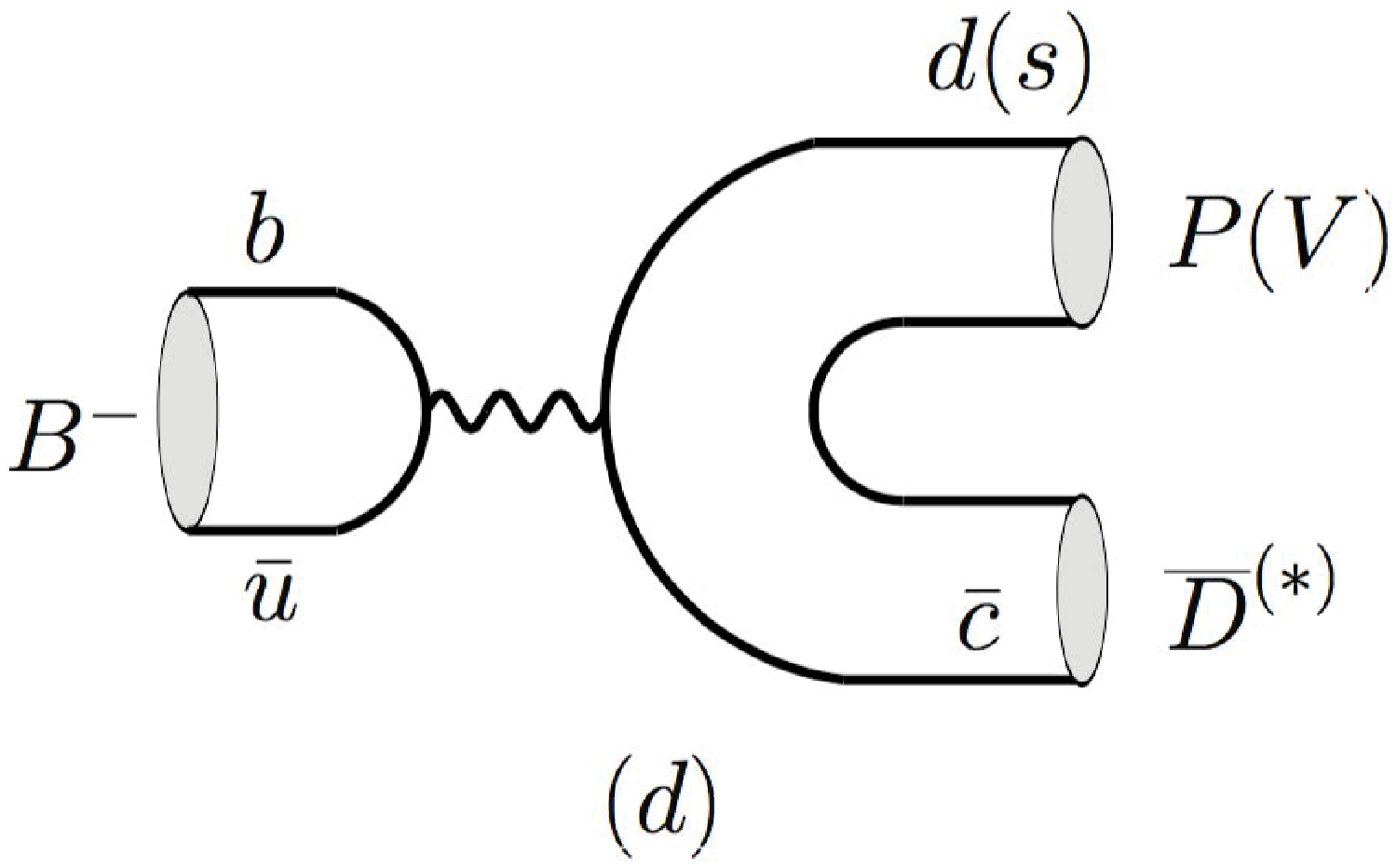}
\end{center}
\caption{Topological diagrams in the $b\to u$ transitions: (a) the color-favored   tree diagram, $T$; (b) the color-suppressed   tree diagram, $C$; (c) the $W$-exchange annihilation-type diagram, $E$; and (d) the $W$-annihilation diagram, $A$. Note that the $E$ diagram occurs only in the $\overline B_{d}^{0}$ and $\overline B_{s}^{0}$ decays, while the $A$ diagram occurs only in $B^{-}$ decays.
}\label{fig:topbu}
\end{figure}

In this section, we shall study the CKM suppressed processes induced by $b\to u \bar c d(s)$ transitions, i.e. $\overline B \to \overline D P$, $\overline D^{*}P$, $\overline DV$ decay modes. The relevant effective Hamiltonian can be obtained by an exchange of $c\leftrightarrow u$ in that of the $b\to c$ transiting processes shown in eq.(\ref{eq2}), as
\begin{equation}
\mathcal{H}_{\rm eff}={G_{F}\over\sqrt2}V_{ub}V_{cq}^{*} \left[C_{1}(\mu)O_{1}(\mu)+C_{2}(\mu)O_{2}(\mu)\right]+h.c.,
\end{equation}
where the two tree-level current-current operators are
\begin{align}
O_{1}=\bar q_{\alpha} \gamma^{\mu}(1-\gamma_{5}) c_{\beta} \bar u_{\beta}\gamma_{\mu}(1-\gamma_{5}) b_{\alpha},
O_{2}=\bar q_{\alpha} \gamma^{\mu}(1-\gamma_{5}) c_{\alpha} \bar u_{\beta}\gamma_{\mu}(1-\gamma_{5}) b_{\beta}.
\end{align}
According the effective Hamiltonian, we can draw the topological diagrams of the $b\to u$ transitions as shown in Fig.\ref{fig:topbu}.

The topologies of the  processes $\overline B \to \overline D^{(*)}M$ induced by $b\to u$ transition are different from $\overline B \to  D^{(*)}M$ induced by $b\to c$. The charmed meson is recoiled in $\overline B \to D^{(*)}M$, while it will be emitted in $\overline B \to \overline D^{(*)}M$ process. It is thus expected that the branching fractions of $\overline B \to \overline D^{(*)}M$ is smaller than those of $\overline B \to  D^{(*)}M$ due to the suppression of CKM elements. The formulae of $\overline B \to \overline D^{(*)}M$ factorizable contributions should be similar to those of $\overline B \to  D^{(*)}M$, but   four new non-factorizable contributions, i.e. $\chi_u^{C,E}$ and $\phi_u^{C,E}$  should be introduced. In principle, these parameters should be extracted from experimental data as done in the $b\to c$ processes, but there are no $C$- or $E$-diagram dominated mode measured in experiments so far. In this case, we shall employ an approximation that the four non-factorizable parameters in the $b\to u$ processes are the same as those in the $b\to c$ processes. Therefore, the formulae of $T$, $C$ and $E$ diagrams are the same in these two kinds of processes, i.e. $\chi_u^{C}= \chi_c^{C}$, $\phi_u^{C}= \phi_c^{C}$, $\chi_u^{E}= \chi_c^{E}$ and $\phi_u^{E}= \phi_c^{E}$. In the following, we will neglect the subscripts of $\chi_{u,c}^{C,E}$ and $\phi_{u,c}^{C,E}$ for simplicity without confusions.

Apart from above contributions, the $W$-annihilation diagram $A$ appears in the $b\to u$ transitions. Again, no experimental data available to fit the contribution of this diagram. Unlike the $E$ diagram dominated by non-factorizable contributions, the factorizable contributions in the $A$ diagram is too large to be neglected. On the contrary, the non-factorizable contributions are suppressed due to the small Wilson coefficient $C_{1}/3$. To calculate the factorizable contribution in the $A$ diagram quantitatively, we will adopt the pole model \cite{Li:2012cfa, Fusheng:2011tw, Li:2013xsa, Kramer:1997yh}, which has been proved to be one effective approach in dealing with the $W$-annihilation diagrams. So, the amplitudes are expressed as
\begin{align}
A_u^{DP}&=-i {G_{F}\over\sqrt2} V_{ub}V_{cq}^{*}\left(C_{2}(\mu)+{C_{1}(\mu)\over N_{c}}\right) f_{B} {f_{D_{0}^{*}} g_{D_{0}^{*}DP} m_{D_{0}^{*}}^{3}\over m_{B}^{2}-m_{D_{0}^{*}}^{2}},
\\
A_u^{D^{*}P}&= \sqrt{2}{G_{F}} V_{ub}V_{cq}^{*}\left(C_{2}(\mu)+{C_{1}(\mu)\over N_{c}}\right) f_{B}{ f_{D} g_{D^{*}DP} m_{D}^{2}\over m_{B}^{2}-m_{D}^{2}}(\varepsilon^{*}_{D^{*}}\cdot p_{B}),
\\
A_u^{DV}&
= \sqrt{2}{G_{F}} V_{ub}V_{cq}^{*}\left(C_{2}(\mu)+{C_{1}(\mu)\over N_{c}}\right) f_{B}  {f_{D} g_{DDV}m_{D}^{2}\over m_{B}^{2}-m_{D}^{2}}(\varepsilon^{*}_{V}\cdot p_{B}),
\end{align}
where the intermediate state is a scalar charmed meson $D_0$ in the $\overline{D}P$ mode, and a pseudoscalar $D$ meson in the $\overline{D}^{*}P$ and $\overline{D}V$ modes. The effective strong coupling constant $g_{D_{0}^{*}DP}=4.2$ is extracted from the experimental data of $D_{0}^{*}(2400)\to D\pi$ decay, $g_{D^{*}DP}=4.8$ is from $D^{*}\to D\pi$ decay \cite{PDG}, and $g_{DDV}=2.52$ is obtained from the vector meson dominance model \cite{Lin:1999ad}. In practice, we will follow the arguments of \cite{Kramer:1997yh} to set all intermediate states on shell, i.e. $p_{\rm pole}^{2}=m_{\rm pole}^{2}$ for simplicity.

With the fitted parameters from processions induced by $b\to c$ transition, the  topological amplitudes and the predicted branching fractions of processes induced by $b\to u$ transition are tabled in Tables~\ref{tab:BrbuDP}, \ref{tab:BrbuDstP} and \ref{tab:BrbuDV}. In these tree tables, the resources of the first three uncertainties are same as processes induced the $b\to c$ transition. Besides, we also include the uncertainties arising from the CKM matrix element $|V_{ub}|$, which has not been well measured till now. From the tables, it is obvious that both form factors and $|V_{ub}|$ take large uncertainties. If the CKM matrix element $|V_{ub}|$ can be determined well, it is expected that the uncertainties from it will be reduced significantly. It should be noted that the decays by $b\to u$ transitions should have additional uncertainties than decay by $b\to c $ transition, since our approximation of same parameters for these two kinds of decays are not well justified. Similarly, we also obtain the hierarchy $|T_u|:|C_u|:|E_u|:|A_u|\sim 1:0.4:0.1:0.03$. Compared with some existed data, our predictions are in agreement with them  with large uncertainties in both sides. And these results will be tested in the LHCb and Belle-II experiments. Note that the branching fractions of the processes induced only by $W$-annihilation diagram are so small that can be regarded as the good place to probe new physics beyond the SM, though these predictions in the current work are somewhat model dependent.

\begin{table}[tbph]
\caption{Branching fractions and  decay amplitudes for the $\overline B\to \overline DP$ modes. Data are from \cite{PDG}. The first uncertainties are from the fitted parameters, the second uncertainties are from the form factors, the third ones are coming from decay constants and the last ones are from $|V_{ub}|$.} \label{tab:BrbuDP}
\begin{center}
\begin{tabular}{lllllll}
\hline\hline
Meson~~~~~~~~ &
Mode~~~~~~~~~ &
Amplitudes~~~~~~~~ &
$\mathcal{B}_{\rm exp}(\times10^{-6})$~~~~~~~~&
$\mathcal{B}_{\rm th}(\times10^{-6})$~~~~~~~~\\
\hline&
Cabibbo-favored &
$V_{ub}V_{cs}^{*}$\\
\hline
$\overline B^{0}$    &
$D_{s}^{-}\pi^{+}$   &
$T$&
$21.6\pm2.6$&
$29.1\pm0.0\pm6.3\pm1.0\pm7.0$\\&
$\overline D^{0}\overline K^{0}$ &
$C$&  &
$5.7\pm0.3\pm1.2\pm 0.3\pm1.4$\\
\hline
$B^{-}$&
$D_{s}^{-}\pi^{0}$ &
${1\over\sqrt2}T$ &
$16\pm5$ &
$15.6\pm0.0\pm3.4\pm0.6\pm3.8$\\&
$D_{s}^{-}\eta$ &
${1\over\sqrt2}T\cos\phi$ &
$<400$ &
$9.8\pm 0.0\pm 2.0\pm 0.3\pm 2.4$\\&
$D_{s}^{-}\eta'$ &
${1\over\sqrt2}T\sin\phi$ & &
$5.9\pm 0.0\pm 1.3\pm 0.2\pm 1.4$\\&
$\overline D^{0}K^{-}$ &
$C+A$ &   &
$5.8\pm0.3\pm 1.3\pm 0.3\pm 1.4$\\ &
$D^{-}\overline K^{0}$ &
$A$ &
$<2.9$ &
$ 0.012\pm 0.000\pm 0.000\pm 0.000\pm 0.003$\\
\hline
$\overline B_{s}^{0}$ &
$D_{s}^{-}K^{+}$ &
$T+E$  &  &
$27.5^{+0.3}_{-0.2}\pm6.6 \pm 1.0 \pm 6.6 $\\ &
$\overline D^{0}\eta$ &
${1\over\sqrt2}E\cos\phi-C\sin\phi$ & &
$ 2.0\pm0.1\pm 0.4\pm 0.1\pm 0.5$\\ &
$\overline D^{0}\eta'$ &
${1\over\sqrt2}E\sin\phi+C\cos\phi$ & &
$ 2.9\pm0.1 \pm 0.6\pm 0.1\pm 0.7$\\&
$D^{-}\pi^{+}$ & $E$ & &
$ 0.14\pm 0.02\pm 0.00\pm 0.02\pm 0.03$\\&
$\overline D^{0}\pi^{0}$ &
${1\over\sqrt2}E$ & &
$0.07\pm 0.01\pm 0.00\pm 0.01\pm 0.02$\\
\hline
\hline
& Cabibbo-suppressed &
$V_{ub}V_{cd}^{*}$\\
\hline
$\overline B^{0}$ &
$D^{-}\pi^{+}$ &
$T+E$ &
$0.78\pm0.14$ &
$0.90\pm 0.01\pm 0.20\pm 0.04\pm 0.22$\\ &
$\overline D^{0}\pi^{0}$ &
${1\over\sqrt2}(E-C)$ &  &
$0.11\pm0.01\pm 0.02\pm 0.01\pm 0.03$\\ &
$\overline D^{0}\eta$ &
${1\over\sqrt2}(E+C)\cos\phi$ &  &
$0.07\pm0.01\pm 0.01\pm 0.00\pm 0.02$\\ &
$\overline D^{0}\eta'$ &
${1\over\sqrt2}(E+C)\sin\phi$ &  &
$0.05\pm0.00\pm 0.01\pm 0.00\pm 0.01$\\ &
$D_{s}^{-}K^{+}$ &
$E$ &  &
$0.011\pm 0.001\pm 0.000\pm 0.001\pm 0.003$\\
\hline
$B^{-}$ &
$\overline D^{0}\pi^{-}$ & $C+A$ &  &
$0.23\pm 0.01\pm 0.05\pm 0.01\pm 0.05$\\ &
$ D^{-}\pi^{0}$ &
${1\over\sqrt2}(T-A)$ &  &
$0.55\pm 0.00\pm 0.12\pm 0.03\pm 0.13$\\ &
$D^{-}\eta$ &
${1\over\sqrt2}(T+A)\cos\phi$ &  &
$0.30\pm 0.00\pm 0.06\pm 0.02\pm 0.07$\\ &
$D^{-}\eta'$ & ${1\over\sqrt2}(T+A)\sin\phi$ &  &
$0.20\pm 0.00\pm 0.04\pm 0.01\pm 0.05$\\ &
$D_{s}^{-}K^{0}$ &
$A$ &
$<800$ &
$ 0.0006\pm 0.0000\pm 0.0000\pm 0.0001\pm 0.0002$\\
\hline
$\overline B_{s}^{0}$ &
$D^{-}K^{+}$ &
$T$ &  &
$1.05\pm 0.00\pm 0.24\pm 0.05\pm 0.25$\\ &
$\overline D^{0}K^{0}$ &
$C$ &&
$0.24\pm 0.01\pm 0.05\pm 0.01\pm 0.06$
\\
\hline\hline
\end{tabular}
\end{center}
\end{table}
\begin{table}[tbph]
\caption{Branching fractions and  decay amplitudes for the $\overline B\to \overline D^{*}P$ decays. }\label{tab:BrbuDstP}
\begin{center}
\begin{tabular}{lllllll}
\hline\hline
Meson~~~~~~~~ &
Mode~~~~~~~~~ &
Amplitudes~~~~~~~~ &
$\mathcal{B}_{\rm exp}(\times10^{-6})$~~~~~~~~&
$\mathcal{B}_{\rm th}(\times10^{-6})$~~~~~~~~\\
\hline &
Cabibbo-favored &
$V_{ub}V_{cs}^{*}$\\ \hline
$\overline B^{0}$ &
$D_{s}^{*-}\pi^{+}$ &
$T$ &
$21\pm4$&
$31.0\pm{0.0}\pm 6.6\pm 3.1 \pm 7.4 $\\ &
$\overline D^{*0}\overline K^{0}$&
$C$&  &
$6.4\pm0.3\pm 1.4\pm 0.6\pm 1.5$\\\hline
$B^{-}$ &
$D_{s}^{*-}\pi^{0}$ &
${1\over\sqrt2}T$ &
$<260$ &
$16.6\pm0.0\pm 3.6\pm 1.7\pm 4.0$\\ &
$D_{s}^{*-}\eta$ &
${1\over\sqrt2}T\cos\phi$ &
$<600$ &
$6.0\pm 0.0\pm 1.6\pm 0.8\pm 1.4$\\ &
$D_{s}^{*-}\eta'$ &
${1\over\sqrt2}T\sin\phi$ & &
$10.7\pm 0.0\pm 1.7\pm 0.9\pm 2.6$\\ &
$\overline D^{*0}K^{-}$ &
$C+A$ &   &
$11.8^{+0.5}_{-0.6}\pm 1.9\pm 0.9\pm 2.8$\\
& $D^{*-}\overline K^{0}$ &
$A$ &
$<9.0$ &
$1.3\pm 0.0\pm 0.0\pm 0.1\pm 0.3$\\\hline
$\overline B_{s}^{0}$ &
$D_{s}^{*-}K^{+}$ &
$T+E$  &  &
$29.7^{+0.3}_{-0.2}\pm 7.1\pm 3.0\pm 7.1$\\
 &
$\overline D^{*0}\eta$ &
${1\over\sqrt2}E\cos\phi-C\sin\phi$ & &
$2.3\pm{ 0.1}\pm 0.5\pm 0.2\pm0.6$\\ &
$\overline D^{*0}\eta'$ &
${1\over\sqrt2}E\sin\phi+C\cos\phi$ & &
$ 3.1\pm0.1\pm 0.6\pm 0.3\pm 0.8$\\ &
$ D^{*-}\pi^{+}$ &
$E$ & &
$ 0.11\pm0.01\pm 0.00\pm 0.02\pm 0.03$\\ &
$\overline D^{*0}\pi^{0}$ &
${1\over\sqrt2}E$ & &
$0.06\pm0.01\pm 0.00\pm 0.01\pm 0.01$\\
\hline\hline
& Cabibbo-suppressed &
$V_{ub}V_{cd}^{*}$\\
\hline
$\overline B^{0}$ &
$D^{*-}\pi^{+}$ &
$T+E$ &  &
$1.0\pm 0.0\pm 0.2\pm 0.1\pm 0.2$\\ &
$\overline D^{*0}\pi^{0}$ &
${1\over\sqrt2}(E-C)$ &  &
$ 0.12\pm 0.01\pm 0.03\pm 0.01\pm 0.03$\\ &
$\overline D^{*0}\eta$ &
${1\over\sqrt2}(E+C)\cos\phi$ &  &
$ 0.08\pm0.0\pm 0.02\pm 0.02\pm 0.02$\\ &
$\overline D^{*0}\eta'$ &
${1\over\sqrt2}(E+C)\sin\phi$ &  &
$ 0.05\pm0.0\pm 0.01\pm 0.00\pm 0.01$\\ &
$D_{s}^{*-}K^{+}$ &
$E$ & &
$ 0.008\pm0.001\pm 0.000\pm 0.001\pm 0.002$\\
\hline
$B^{-}$ &
$\overline D^{*0}\pi^{-}$ &
$C+A$ &  &
$0.43\pm 0.02\pm 0.07\pm 0.03\pm 0.10$\\ &
$ D^{*-}\pi^{0}$ &
${1\over\sqrt2}(T-A)$ &
$<3.6$  &
$ 0.40\pm 0.00\pm 0.11\pm 0.05\pm 0.10$\\ &
$ D^{*-}\eta$ &
${1\over\sqrt2}(T+A)\cos\phi$ &  &
$0.48\pm 0.00\pm 0.09\pm 0.04\pm 0.12$\\ &
$ D^{*-}\eta'$ &
${1\over\sqrt2}(T+A)\sin\phi$ &  &
$ 0.31\pm 0.00\pm 0.06\pm 0.03\pm 0.07$\\ &
$ D_{s}^{*-}K^{0}$ &
$A$ &
$<900$ &
$ 0.03\pm0.00\pm 0.00\pm 0.00\pm 0.01$\\
\hline
$\overline B_{s}^{0}$ &
$D^{*-}K^{+}$ &
$T$ &  &
$ 1.17\pm 0.00\pm 0.27\pm 0.12\pm 0.28$\\ &
$\overline D^{*0}K^{0}$ &
$C$ &  &
$ 0.27\pm0.01\pm 0.06\pm 0.03\pm 0.06$\\
\hline\hline
\end{tabular}
\end{center}
\end{table}
\begin{table}[tbph]
\caption{Branching fractions and  decay amplitudes for the $\overline B\to \overline DV$ decays. }\label{tab:BrbuDV}
\begin{center}
\begin{tabular}{lllllll}
\hline\hline
Meson~~~~~~~~ &
Mode~~~~~~~~~ &
Amplitudes~~~~~~~~ &
$\mathcal{B}_{\rm exp}(\times10^{-6})$~~~~~~~~&
$\mathcal{B}_{\rm th}(\times10^{-6})$~~~~~~~~\\
\hline
& Cabibbo-favored &
$V_{ub}V_{cs}^{*}$\\
\hline
$\overline B^{0}$ &
$D_{s}^{-}\rho^{+}$ &
$T$ &
$<24$  &
$31.2\pm 0.0  \pm  7.5\pm 1.1 \pm 7.5$\\  &
$\overline D^{0}\overline K^{*0}$ &
$C$ &
$<11$ &
$5.2\pm0.3\pm 1.2\pm 0.2\pm 1.2$\\
\hline
$B^{-}$  &
$D_{s}^{-}\rho^{0}$ &
${1\over\sqrt2}T$ &
$<300$ &
$16.8\pm 0.0 \pm 4.0\pm 0.6\pm 4.0$\\ &
$D_{s}^{-}\omega$ & ${1\over\sqrt2}T$ &
$<400$ &
$12.7 \pm 0.0 \pm 3.1\pm 0.5\pm 3.1$\\ &
$\overline D^{0}K^{*-}$ &
$C+A$ &   &
$ 11.2^{+ 0.5}_{- 0.6}\pm 1.7\pm 0.5\pm 2.7$\\ &
$D^{-}\overline K^{*0}$ &
$A$ &
$<1.8$ &
$1.8\pm 0.0\pm 0.0\pm 0.2\pm 0.4$\\  &
$D_{s}^{-}\phi$ &
$A$ &
$1.7^{+1.2}_{-0.7}$ &
$1.2\pm0.0\pm 0.0\pm 0.1\pm 0.3$\\
\hline
$\overline B_{s}^{0}$ &
$D_{s}^{-}K^{*+}$ &
$T+E$  & &
$22.4^{+0.3}_{-0.2}\pm4.3\pm 0.8\pm 5.4$\\  &
$\overline D^{0}\phi$ &
$C$ & &
$4.4\pm0.2\pm 1.1\pm 0.2\pm 1.1$\\ &
$D^{-}\rho^{+}$ &
$E$ & &
$0.25^{+ 0.03}_{- 0.01}\pm 0.00\pm 0.04\pm 0.06$\\ &
$\overline D^{0}\rho^{0}$ &
${1\over\sqrt2}E$ & &
$0.13^{+ 0.02}_{- 0.01}\pm 0.00\pm 0.02\pm 0.03$\\ &
$\overline D^{0}\omega$ &
${1\over\sqrt2}E$ & &
$0.11\pm0.01\pm 0.00\pm 0.02\pm 0.03$\\
\hline
\hline
&Cabibbo-suppressed &
$V_{ub}V_{cd}^{*}$\\\hline
$\overline B^{0}$ &
$D^{-}\rho^{+}$ &
$T+E$ &   &
$0.94\pm 0.01\pm 0.24\pm 0.05\pm 0.22$\\&
$\overline D^{0}\rho^{0}$ &
${1\over\sqrt2}(E-C)$ &  &
$0.12\pm0.01\pm 0.03\pm 0.01\pm 0.03$\\ &
$\overline D^{0}\omega$ &
${1\over\sqrt2}(E+C)$ &  &
$0.10\pm0.01\pm 0.02\pm 0.01\pm 0.02$\\&
$D_{s}^{-}K^{*+}$ &
$E$ & &
$0.014^{+ 0.002}_{- 0.001}\pm 0.000\pm 0.002\pm 0.003$\\
\hline
$B^{-}$  &
$ D^{-}\rho^{0}$ &
${1\over\sqrt2}(T-A)$ &  &
$ 0.33\pm 0.00\pm 0.10\pm 0.02\pm 0.08$\\ &
$ D^{-}\omega$ &
${1\over\sqrt2}(T+A)$ &  &
$ 0.69\pm0.00\pm 0.13\pm 0.04\pm 0.17$\\&
$\overline D^{0}\rho^{-}$ &
$C+A$ &  &
$0.48\pm 0.02\pm 0.08\pm 0.02\pm 0.11$\\ &
$ D_{s}^{-}K^{*0}$ &
$A$ &
$<4.4$ &
$0.04\pm 0.00\pm 0.00\pm 0.01\pm 0.01$\\
\hline
$\overline B_{s}^{0}$ &
$D^{-}K^{*+}$ &
$T$ &  &
$0.88\pm 0.00\pm 0.16\pm 0.04\pm 0.21$\\ &
$\overline D^{0}K^{*0}$ &
$C$ & &
$0.20\pm 0.01\pm 0.04\pm 0.01\pm 0.05$\\
\hline\hline
\end{tabular}
\end{center}
\end{table}

\section{phenomenological Implications} \label{sec:4}
In this section, we are going to discuss the isospin asymmetry, SU(3) and $CP$ asymmetry in turn.
\subsection{Isospin Analysis}
The $\overline B\to D \pi$ system can   be decomposed in terms of two isospin amplitudes, $A_{1/2}$ and $A_{3/2}$, which correspond to the transition into $D\pi$ final states with isospin $I=1/2$ and $I=3/2$, respectively. In the experimental side, the ratio
\begin{equation}\label{iso-rat}
\frac{A_{1/2}}{\sqrt{2} A_{3/2}}=1+{\cal O}(\Lambda_{\rm QCD}/m_b)
\end{equation}
is a measure of the departure from the heavy-quark limit~\cite{Beneke:2000wa}. The corresponding isospin relations read as
\begin{subequations}
\begin{align}
A(\overline B^0_d\to D^+\pi^-)& =\sqrt{\frac{1}{3}}A_{3/2}+\sqrt{\frac{2}{3}}A_{1/2}=T+E,\\
\sqrt{2}A(\overline B^0_d\to D^0\pi^0)& =\sqrt{\frac{4}{3}}A_{3/2}-\sqrt{\frac{2}{3}}A_{1/2}=C-E,\\
A(B_u^- \to D^0\pi^-)&=\sqrt{3}A_{3/2}=T+C .
\end{align}
\end{subequations}
So the isospin amplitudes can be expressed by the topological amplitudes as
\begin{align}
A_{1/2}=\frac{2T-C+3E}{\sqrt{6}},~~~~~~~~~~
A_{3/2}=\frac{T+C}{\sqrt{3}},
\label{eq:a32}
\end{align}
which leads to the following expression,
\begin{equation}\label{iso-rat-calc}
\frac{A_{1/2}}{\sqrt{2}A_{3/2}}=1-\frac{3}{2}\left(\frac{C-E}{T+C} \right).
\end{equation}
The relative strong phase between the $I=3/2$ and $I=1/2$ amplitudes can be calculated with
\begin{equation}
\cos \delta = \frac{3|A(D^+\pi^-)|^2 + |A(D^0\pi^-)|^2 - 6 |A(D^0\pi^0)|^2}{6
\sqrt{2}|A_{1/2}||A_{3/2}|}.
\end{equation}
In this work, we find the following numerical results
\begin{eqnarray}
\left|\frac{A_{1/2}}{\sqrt{2}A_{3/2}}\right|_{D\pi}= 0.65 \pm 0.03,
\end{eqnarray}
which are complemented by
\begin{eqnarray}
\cos \delta= 0.90\pm 0.04 .
\end{eqnarray}
The corresponding central values for the strong phases then become $\delta = 25^\circ$. Comparing with eq.(\ref{iso-rat}), we observe that the isospin-amplitude ratio shows significant deviation from the heavy-quark limit. Because the  contribution from annihilations has been neglected,  we can trace this feature back to the large color-suppressed $C$ topologies.
\subsection{SU(3) Symmetry Breaking}
Now we turn to discuss the SU(3) symmetry breaking effect in the charmed $B$ decays. If flavor  SU(3) were exact, one would get
\begin{itemize}
  \item For $\overline B\to DK$, $\overline B \to D\pi$, $\overline B_s\to D_sK$ and $ \overline B_s\to D_s\pi$
  \begin{subequations}
  \begin{align}
  &\left|\frac{T^{\overline B\to DK}}{V_{cb}V_{us}^*}\right|
  =\left|\frac{T^{\overline B\to D\pi}}{V_{cb}V_{ud}^*}\right|
  =\left|\frac{T^{\overline B_s\to D_sK}}{V_{cb}V_{us}^*}\right|
  =\left|\frac{T^{\overline B_s\to D_s\pi}}{V_{cb}V_{ud}^*}\right|,\\
  &\left|\frac{C^{\overline B\to DK}}{V_{cb}V_{us}^*}\right|
  =\left|\frac{C^{\overline B\to D\pi}}{V_{cb}V_{ud}^*}\right|
  =\left|\frac{C^{\overline B_s \to DK}}{V_{cb}V_{us}^*}\right|;
  \end{align}
  \end{subequations}
  \item For $\overline B\to DK^{*}$, $\overline B\to D\rho$, $\overline B_s\to D_sK^{*}$ and $\overline B_s\to D_s\rho$
  \begin{subequations}
  \begin{align}
  &\left|\frac{T^{\overline B\to DK^{*}}}{V_{cb}V_{us}^*}\right|
  =\left|\frac{T^{\overline B\to D\rho}}{V_{cb}V_{ud}^*}\right|
  =\left|\frac{T^{\overline B_s \to D_sK^{*}}}{V_{cb}V_{us}^*}\right|
  =\left|\frac{T^{\overline B_s \to D_s\rho}}{V_{cb}V_{ud}^*}\right|,\\
  &\left|\frac{C^{\overline B\to DK^{*}}}{V_{cb}V_{us}^*}\right|
  =\left|\frac{C^{\overline B\to D\rho}}{V_{cb}V_{ud}^*}\right|
  =\left|\frac{C^{\overline B_s \to DK^{*}}}{V_{cb}V_{us}^*}\right|;
  \end{align}
  \end{subequations}
   \item For the annihilation type decay modes $\overline B\to D_sK^{(*)}$ and  $\overline B_s\to D\pi(\rho)$
   \begin{subequations}
  \begin{align}
  &\left|\frac{E^{\overline B\to D_sK}}{V_{cb}V_{ud}^*}\right|
  =\left|\frac{E^{\overline B_s^0\to D\pi}}{V_{cb}V_{us}^*}\right|,\\
  &\left|\frac{E^{\overline B\to D_sK^*}}{V_{cb}V_{ud}^*}\right|
  =\left|\frac{E^{\overline B_s^0\to D \rho}}{V_{cb}V_{us}^*}\right|.
  \end{align}
  \end{subequations}
\end{itemize}
To estimate the SU(3) breaking effect, we use the $\chi^2$ fit results and obtain
\begin{subequations}
\begin{align}
   \left|\frac{T^{\overline B\to DK}}{V_{cb}V_{us}^*}\right|
  :\left|\frac{T^{\overline B\to D\pi}}{V_{cb}V_{ud}^*}\right|
  :\left|\frac{T^{\overline B_s\to D_sK}}{V_{cb}V_{us}^*}\right|
  :\left|\frac{T^{\overline B_s\to D_s\pi}}{V_{cb}V_{ud}^*}\right|&=1:0.83:1.10:0.90;\\
   \left|\frac{C^{\overline B\to DK}}{V_{cb}V_{us}^*}\right|
  :\left|\frac{C^{\overline B\to D\pi}}{V_{cb}V_{ud}^*}\right|
  :\left|\frac{C^{\overline B_s \to DK}}{V_{cb}V_{ud}^*}\right|&=1:0.85:0.91;\\
   \left|\frac{T^{\overline B\to DK^{*}}}{V_{cb}V_{us}^*}\right|
  :\left|\frac{T^{\overline B\to D\rho}}{V_{cb}V_{ud}^*}\right|
  :\left|\frac{T^{\overline B_s \to D_sK^{*}}}{V_{cb}V_{us}^*}\right|
  :\left|\frac{T^{\overline B_s \to D_s\rho}}{V_{cb}V_{ud}^*}\right|&=1:0.83:1.07:0.89;\\
   \left|\frac{C^{\overline B\to DK^{*}}}{V_{cb}V_{us}^*}\right|
  :\left|\frac{C^{\overline B\to D\rho}}{V_{cb}V_{ud}^*}\right|
  :\left|\frac{C^{\overline B_s \to DK^{*}}}{V_{cb}V_{ud}^*}\right|&=1:0.79:0.84;\\
     \left|\frac{E^{\overline B\to D_sK}}{V_{cb}V_{ud}^*}\right|
  :\left|\frac{E^{\overline B_s\to D  \pi}}{V_{cb}V_{us}^*}\right|&=1:0.81;\\
   \left|\frac{E^{\overline B\to D_sK^*}}{V_{cb}V_{ud}^*}\right|
  :\left|\frac{E^{\overline B_s^0\to D\rho}}{V_{cb}V_{us}^*}\right|&=1:0.80.
  \end{align}
 \end{subequations}
The above results show that the SU(3) symmetry breaking in $\overline B \to DM$  is about $10\sim20\%$ at the amplitude level.

Now, let us look at the SU(3) symmetry breaking in the $B_u^-\to D^0 K^-$ and $B_u^-\to D^0\pi^-$, which are related by the so-called U-spin symmetry.  For the amplitudes, both $T$ and $C$ topologies contribute to them, and $T$ is proportional to the decay constant of light meson, while $C$ is proportional to the form factor of $B$ to light meson. Due to a good approximation $F_{0}^{B\to K}/F_{0}^{B\to\pi}\approx f_{K}/ f_{\pi}$, we then obtain the ratio of the above two processes
\begin{equation}
\mathcal{R}_{1}=
\frac{\mathcal{B}(B_u^{-}\to D^{0}K^{-})/|V_{us}f_{K}|^{2}}{\mathcal{B}(B_u^{-}\to D^{0}\pi^{-})/|V_{ud}f_{\pi}|^{2}} = 1.00,
\end{equation}
which agrees well with the experimental data
\begin{equation}
\mathcal{R}_{1}^{\rm exp} = 1.005\pm0.056.
\end{equation}
Thus, we conclude that for decay modes dominated by $T$ terms, the source of SU(3) symmetry breaking is mainly from the decay constants of light mesons.

In addition, the combination of  decay modes $\overline B_s^0\to D_s^{*\mp} K^\pm$ and $\overline B_s^0\to D_s^{*\mp} \pi^\pm$\footnote{In fact, the decays $\overline B_s^0\to D_s^{(*)-} \pi^+$ do not exist, and we write them here for symmetrization.} is used to test SU(3) symmetry, and the ratios between $\overline B_{s}\to D_{s}^{(*)\mp}K^{\pm}$ and $\overline B_{s}\to D_{s}^{\mp}\pi^{\pm}$ is given by
\begin{align}
&\mathcal{R}_2\equiv {\mathcal{B}(\overline B_{s}^{0}\to D_{s}^{\mp}K^{\pm})\over \mathcal{B}(\overline B_{s}^{0}\to D_{s}^{\mp}\pi^{\pm})},\,\,\,\,\,\,
\mathcal{R}_2^{*}\equiv {\mathcal{B}(\overline B_{s}^{0}\to D_{s}^{*\mp}K^{\pm})\over \mathcal{B}(\overline B_{s}^{0}\to D_{s}^{*\mp}\pi^{\pm})}
\end{align}
Under SU(3) limit, the two ratios are given by \cite{DeBruyn:2012jp}
\begin{align}
&\mathcal{R}_2|_{\rm SU(3)}\ =0.0864^{+0.0087}_{-0.0072}, \,\,\,\,\,\,\,
\mathcal{R}_2^*|_{\rm SU(3)}=0.099^{+0.030}_{-0.036}.\label{eq:r2su3}
\end{align}
The results we obtained are:
\begin{align}
&\mathcal{R}_2|_{\rm FAT}=0.079^{+0.013}_{-0.005},\,\,\,\,\,\,\,
\mathcal{R}_2^*|_{\rm FAT}=0.081^{+0.005}_{-0.003}.\label{eq:r2fat}
\end{align}
Very recently, LHCb published the latest results on these two ratios\cite{Aaij:2014jpa,Aaij:2015dsa}:
\begin{align}
&\mathcal{R}_2|_{\rm Exp}=0.0762\pm0.0015\pm0.0020,\,\,\,\,\,\,\,
\mathcal{R}_2^*|_{\rm Exp}=0.068\pm0.005^{+0.003}_{-0.002}.\label{eq:r2exp}
\end{align}
Comparing results of Eq.(\ref{eq:r2su3}), Eq.(\ref{eq:r2fat}), and (\ref{eq:r2exp}), it is obvious that our result  for $R_2$ falls into the range between SU(3) limit and experimental data, while  for $R_2^*$ both theoretical predictions are larger than the data, which implies that the SU(3) symmetry breaking effect might be more sizable than we expected in these two decays.

\subsection{CP Asymmetry}
\begin{figure}[htb]
\begin{center}
\includegraphics[scale=0.6]{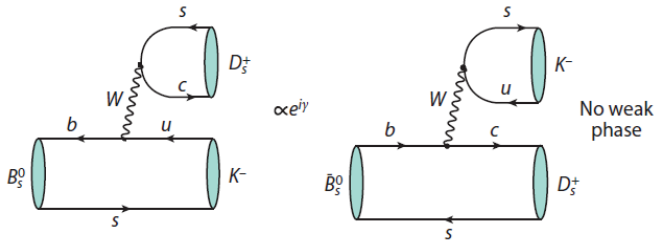}
\end{center}
\caption{Feynman diagrams of $B_s\to D_s^\pm K^\mp$}\label{fig:BsDsK}
\end{figure}

Among $B_s$ decays, special attention is paid to the decay modes $\overline B_s\to D_s^\pm K^\mp$. As shown in Figure \ref{fig:BsDsK}, $B_s^0(\overline B_s^0)\to D_s^\pm K^\mp$ decays receive contributions only from $T$ topological amplitudes; in other words, there are no penguin contributions. Note that both $B_s^0$ and $\overline B_s^0$ mesons can decay into the $D_s^+K^-$ final state via CKM matrix elements $V_{ub}V_{cs}$ and $V_{cb}V_{us}$, respectively.     They   are both of the same order, $\lambda^3$, in the Wolfenstein expansion, allowing for large interference effects. Consequently, interference effects between $B_s^0-\overline B_s^0$ mixing and decay processes lead to a time-dependent CP asymmetry, which provides sufficient information to determine the weak phase $\gamma$ in a theoretically clean way. In the following discussion, we set $f=D_s^-K^+$ for simplicity. The time-dependent decay rates of the initially produced flavor eigenstates $|B^0_s(t = 0)\rangle$ and $|\overline B_s^0(t=0)\rangle$ are given by \cite{LHCb:2012vja}
\begin{subequations}
\begin{eqnarray}
&\frac{d\Gamma(B_s^0(t)\rightarrow f)}{dt} = \frac{1}{2}|A_f|^2 e^{-\Gamma t} (1+|\lambda_f|^2) \left\{ \cosh(\frac{\Delta\Gamma  t}{2})-D_f\sinh(\frac{\Delta\Gamma  t}{2})
+C_f\cos(\Delta m_s t)-S_f\sin(\Delta m_s t)\right\} , \\
&\frac{d\Gamma(\overline B_s^0(t)\rightarrow f)}{dt} = \frac{1}{2}|A_f|^2 (\frac{p}{q})^2 e^{-\Gamma t} (1+|\lambda_f|^2)  \left\{ \cosh(\frac{\Delta\Gamma  t}{2})-D_f\sinh(\frac{\Delta\Gamma  t}{2})
-C_f\cos(\Delta m_s t)+S_f\sin(\Delta m_s t)\right\} ,
\end{eqnarray}
\end{subequations}
where  $A_f$ is the amplitude of  $B_s^0 \to f$, and the definition  of $\lambda_f$ is
\begin{eqnarray}
\lambda_f = \frac{q}{p}\frac{\bar A_f}{A_f}=\frac{q}{p} \frac{A(\overline B_s^0 \rightarrow f)}{A(B_s^0\rightarrow f)}.
\label{equ:lamdaf}
\end{eqnarray}
The complex coefficients $p$ and $q$ relate the $B^0_s$ meson mass eigenstates $|B_{H,L}\rangle$ to the flavor eigenstates  $B_s^0$ and $\overline B_s^0$,
\begin{align}
|B_L\rangle  = p|B_s^0\rangle + q|\overline B_s^0 \rangle , \qquad
|B_H\rangle  = p|B_s^0\rangle - q|\overline B_s^0 \rangle,
\end{align}
and $|p|^2+|q|^2=1$ is satisfied. In the Standard Model, $q/p$ is given by
\begin{eqnarray}
\frac{q}{p} \approx \frac{V_{ts}V_{tb}^*}{V_{ts}^*V_{tb}}\approx e^{-2i\beta_s}.
\end{eqnarray}
Moreover, $\Delta m_s$ and $\Delta \Gamma$ denote the mass difference and the total decay width difference of $B_H$ and $B_L$, respectively. Similar equations can be written for the CP-conjugate decays replacing $A_f$ by $\bar A_{\bar f}=\langle\bar f|\overline B_s^0\rangle$, $\lambda_f$ by $\bar \lambda_{\bar f}=(p/q)(A_{\bar f}/\bar A_{\bar f})$,  $|p/q|^2$ by $|q/p|^2$, $C_f$ by $C_{\bar f}$, $S_f$ by $S_{\bar f}$, and $D_f$ by $D_{\bar f}$. The $CP$ violation parameters are expressed as \cite{PDG}
\begin{align}
&C_f =C_{\bar f}= \frac{1-|\lambda_f|^2}{1+|\lambda_f|^2} , \\
&S_f =\frac{2\mathrm{Im}(\lambda_f)}{1+|\lambda_f|^2} ,\qquad
S_{\bar f} =\frac{2\mathrm{Im}(\bar \lambda_{\bar f})}{1+|\bar \lambda_{\bar f}|^2} ,\\
&D_f =\frac{2\mathrm{Re}(\lambda_f)}{1+|\lambda_f|^2},\qquad
D_{\bar f} =\frac{2\mathrm{Re}(\bar \lambda_{\bar f})}{1+|\bar \lambda_{\bar f}|^2} .
\label{equ:cpobdf}
\end{align}
Note that the equality $C_f =C_{\bar f}$ results from $|q/p| = 1$ and $\lambda_f=\bar \lambda_{\bar f}$. If the above five experimental observables can be measured well and $\beta_s$ can be measured elsewhere, the CKM angle $\gamma$ can be extracted.   The $B_s$ mixing phase $\beta_s$ is predicted to be small in the Standard Model \cite{Lenz:2011ti}, thus we set $\beta_s=-2.5^\circ$ in this work. With the $\chi^2$ fitted result, we then have
\begin{align}
C_f=C_{\bar f}= 0.71\pm0.07 , \,
S_f=-S_{\bar f} =-0.63\pm0.06 , \,
D_f=D_{\bar f} =0.32\pm0.03, \label{equ:cpobth}
\end{align}
where the only uncertainties come from the form factors. In 2011, using a dataset corresponding to $1.0 \mathrm{fb}^{-1}$ recorded in $pp$ collisions at $\sqrt{s}=7 \mathrm{TeV}$, LHCb found the $CP$-violation observables to be \cite{LHCb:2012vja}
\begin{align}
&C_f =1.01\pm0.50\pm0.23, \nonumber\\
&S_f =-1.25\pm0.56\pm0.24,\, S_{\bar f} =0.08\pm0.68\pm 0.28, \nonumber\\
&D_f=-1.33\pm0.60\pm0.26,\,D_{\bar f} =-0.81\pm0.56\pm 0.26, \label{equ:cpobex}
\end{align}
where the first uncertainties are statistical and the second uncertainties are systematic. Comparing our predictions (Eq.(\ref{equ:cpobth})) with the experimental results (Eq.(\ref{equ:cpobex})), we find that our results agree with data within uncertainties. It is should be noted that in our calculation, the $|V_{ub}|$ we used is the averaged value of inclusive and exclusive results. However, there is a clear tension between the $|V_{ub}|$ values extracted from the analysis of inclusive and exclusive decays at present, which may lead to large uncertainties in the theoretical calculations.

In addition, the direct $CP$ asymmetries of $B_s^0 \to D_s^{(*)\pm}K^\mp$ decays are given by \cite{Nandi:2008rg}:
\begin{align}
{\cal A}_{CP}^{(*)}\equiv\frac{
\mathcal{B}(B_{s}^0\to D_{s}^{(*)+}K^{-})+
\mathcal{B}(\overline B_{s}^0\to D_{s}^{(*)+}K^{-})-
\mathcal{B}(B_{s}^0\to D_{s}^{(*)-}K^{+})-
\mathcal{B}(\overline B_{s}^0\to D_{s}^{(*)-}K^{+})
}{\mathcal{B}(B_{s}^0\to D_{s}^{(*)+}K^{-})+
\mathcal{B}(\overline B_{s}^0\to D_{s}^{(*)+}K^{-})+
\mathcal{B}(B_{s}^0\to D_{s}^{(*)-}K^{+})+
\mathcal{B}(\overline B_{s}^0\to D_{s}^{(*)-}K^{+})}.
\end{align}
In this work, because we set $\chi^C_c=\chi^C_u$ and $\phi^C_c=\phi^C_u$ and ignore the life difference between $B_s^0$ and $\overline B_s^0$, we then get:
\begin{align}
{\cal A}_{CP}^{(*)}|_{\rm FAT}=0,
\end{align}
which agree with the predictions considering the life diffrence \cite{DeBruyn:2012jp}
\begin{align}
{\cal A}_{CP}|_{\rm SU(3)}=-0.027^{+0.052}_{-0.019},\,\,\,\,\,\,\,
{\cal A}_{CP}^{*}|_{\rm SU(3)}=-0.035^{+0.056}_{-0.024}.
\end{align}
So, if in future the direct $CP$ asymmetries can be measured at the level of more than ten percent, it would be useful to place tighter bounds on the relation between $\chi_c^C e^{i \phi_c^C}$ and $\chi_u^C e^{i \phi_u^C}$.

\section{Summary}\label{sec:5}
In the work, we preformed analysis of two-body charmed $B$ decays globally using the factorization-assisted topological-amplitude approach. Since the color favored tree emission diagram has been proved factorization in all orders of $\alpha_s$ expansion, we use the factorization results of short-distance Wilson coefficients times the decay constant and form factor. For the color-suppressed tree emission and $W$ exchange diagrams, four universal nonperturbative parameters were introduced, namely $\chi^C$, $\phi^C$, $\chi^E$ and $\phi^E$, the numerical values of which were fitted from the 31 well measured branching fractions. With the fitted results, we then predicted the branching fractions of all 120 $B_{u,d,s}\to D^{(*)}P(V)$ decay modes. For the modes induced by $b\to c$ transition, most results agree with the experimental data well. The number of free parameters and the $\chi^2$ per degree of freedom are both reduced comparing with previous topological diagram analysis.  Due to the suppression by CKM element $|V_{ub}|$, the branching fractions of the processes dominated by $b\to u$ transition are in particular small. Most decays will be measured in the ongoing LHCb experiment and the forthcoming Bell-II experiment. We also found that the SU(3) symmetry breaking is more than $10\%$, and even reach $31\%$ at the amplitude level. For the decays $\overline B_s^0 (B_s^0)\to D_s^\pm K^\mp$, the $CP$ asymmetries predicted agree with data within uncertainty.

\section*{Acknowledgement}
The work is supported by National Natural Science Foundation of China (11175151, 11575151, 11375208, 11521505, 11347027, 11505083 and 11235005) and the Program for New Century Excellent Talents in University (NCET) by Ministry of Education of P. R. China (NCET-13-0991).


\end{document}